\documentclass[aps,prd,11pt,a4paper,nofootinbib,oneside,superscriptaddress]{revtex4-2}

\usepackage{amsmath,amssymb,amsfonts,color}
\usepackage{tensor,slashed,paralist,cases,mathrsfs}
\usepackage{float,cancel,xcolor}
\usepackage{graphicx}
\usepackage{dcolumn}
\usepackage{bm}
\usepackage{tabularx}
\usepackage{multirow}
\usepackage{standalone}
\usepackage[colorlinks=true,urlcolor=blue,linkcolor=blue,citecolor=blue,linktocpage=true]{hyperref}

\newcommand{\beq}{\begin{eqnarray}}
\newcommand{\eeq}{\end{eqnarray}}
\newcommand{\bpmatrix}{\begin{pmatrix}}
\newcommand{\epmatrix}{\end{pmatrix}}
\newcommand{\ba}{\begin{array}}
\newcommand{\ea}{\end{array}}

\renewcommand{\eqref}[1]{Eq.~(\ref{#1})}

\newcommand{\bc}{\begin{center}}
\newcommand{\ec}{\end{center}}

\begin{document}

\vspace*{1.5em}

\title{ Self-Interacting Forbidden Dark Matter under a Cannibally Co-Decaying Phase}

\author{Kwei-Chou Yang}
\email{kcyang@cycu.edu.tw}

\affiliation{Department of Physics and Center for High Energy Physics, Chung Yuan Christian University,
200 Chung Pei Road, Taoyuan 32023, Taiwan}


\begin{abstract}

In a usual dark matter (DM) model without a huge mass difference between the DM and lighter mediator, using the coupling strength suitable for having the correct relic density, the resulting self-interaction becomes several orders of magnitude smaller than that required to interpret the small-scale structures. We present a framework that can offer a solution for this point. We consider a model that contains the vector DM and a heavier but unstable Higgs-like scalar in the hidden sector. 
When the temperature drops below $\sim m_{\rm DM}$, the hidden sector, which is thermally decoupled from the visible sector, enters a cannibal phase, during which the DM density is depleted with the out-of-equilibrium decay of the scalar. The favored parameter region, giving the correct relic density and the proper size of self-interactions, shows the scalar-to-DM mass ratio $\in [1.1,1.33]$ and the scalar mass $\in[9,114]\,{\rm MeV}$.  A sizable parameter space still survives the most current constraints and can be further probed by the near future NA62 beam dump experiment.

\end{abstract}
\maketitle
\newpage

\section{Introduction}

While astronomical observations have strongly indicated dark matter (DM), its identity remains an open question in physics. We thus need to extend the Standard Model (SM) by including new particle(s) to account for the observed DM phenomena.
However, the weak-scale DM search related to its interactions with the SM is stringently constrained by direct detection experiments, especially in the mass range $m_{\rm DM}\gtrsim 5$~GeV \cite{XENON:2018voc, PandaX-4T:2021bab, XENON:2019rxp, Akerib:2018lyp}. This has led to some studies on sub-GeV DM or resorted to some workable models in recent years.
On the other hand, one exciting paradigm of models is the hidden sector DM model, for which the dark matter candidate resides within a hidden sector and communicates with the visible sector through a metastable mediator weakly coupling DM to the SM \cite{Pospelov:2007mp, Berlin:2016gtr}. This model can easily interpret the present experimental null measurements.

One particular class of models that generates the relic density quite differently is called the forbidden dark matter.  For the thermal freeze-out DM of this class, its relic abundance is determined exclusively by forbidden channels, which are kinetically forbidden at zero temperature \cite{Griest:1990kh, Tulin:2012uq, Jackson:2013pjq, Jackson:2013tca, Delgado:2016umt, DAgnolo:2015ujb, DAgnolo:2020mpt, Wojcik:2021xki, Herms:2022nhd, Yang:2022zlh, Liu:2023kat, Aboubrahim:2023yag, Cheng:2023hzw}.
 In this paper, we will assume that the hidden sector, as the SM, contains a Higgs-like scalar $S$. The hidden sector interacts with the SM by mixing the SM-like Higgs and hidden Higgs. We study the simplest scenario where the DM ($X$) is an abelian gauge vector boson but with mass $m_X < m_S$, which resides in the same hidden sector as $S$.  As characterized by the forbidden channel, the forbidden rate of $XX \to SS$, compared with its inverse process at their temperature $T_X$, is exponentially suppressed as
 \begin{align}
\langle \sigma v \rangle_{XX\to SS}
 &= \langle \sigma v \rangle_{SS\to XX} \frac{(n_S^{\text{eq}}(T_X))^2}{(n_X^{\text{eq}}(T_X))^2} \nonumber \\
 & \approx \langle \sigma v \rangle_{SS\to XX} \frac{g_S^2}{g_X^2} r^3
     \bigg(1- \frac{15 \Delta }{8r x_X^2}  \bigg) e^{-\Delta} \,,
\end{align}
where the detailed balance is used, $\Delta \equiv 2(r-1)x_X$, $x_X\equiv m_X/T_X$, the mass ratio $r\equiv m_S/m_X >1$, and
$n_i^{\text{eq}}$ and $g_i$ are the equilibrium number density and internal degrees of freedom (DoFs) of $i $, respectively, with $i \equiv X, S$.

 We aim to find relevant regions of the parameter space in this simple model, where not only the correct relic density can be obtained but also a proper size of DM self-interactions can account for the small scale problems  \cite{Dave:2000ar, Vogelsberger:2012ku, Rocha:2012jg, Peter:2012jh, Zavala:2012us}. We find that, to achieve this purpose, the coupling ($g_{\rm dm}$) between $X$ and $S$ needs to be ${\cal O}(1)$.
 Meanwhile, a much larger mass ratio $m_S/m_X \in (1.5,1.65)$  is required if the hidden sector is maintained in chemical equilibrium with the SM before freezing out.
 However, to maintain such an equilibrium, the mixing angle ($\alpha$) between the hidden Higgs and SM-like Higgs also needs to be much larger, which is nevertheless excluded by the fixed target and flavor experiments (see the discussions in Sec.~\ref{sec:boltz} and Appendix~\ref{thermal_freezeout}).
 
Fortunately, the region of the parameter space with a smaller $\alpha$, resulting in the hidden sector being thermally decoupled from the SM earlier, can have $g_{\rm dm} \propto {\cal O}(1)$ which provides the correct DM relic density and a proper size of DM self-interactions. We will show that the relevant parameter region evades the current constraints and can be further probed in further experiments.

In this relevant scenario, the hidden sector is thermally decoupled from the SM bath when the temperature drops below $\sim m_{\rm DM}$.  After that, it enters a cannibal phase (with zero chemical potential) during which the DM density is heated but depleted with the out-of-equilibrium decay of the mediator; the former describes the number-changing interactions, called cannibalization \cite{Carlson:1992fn, Farina:2016llk}, while the latter is referred to as the co-decaying mechanism \cite{Dror:2016rxc} resulting in the bath reheated. This scenario naturally leads to large self-interactions in the ``sub-GeV" region, which may help address the ‘core vs. cusp’ and ‘too-big-to-fail’ small structure problems \cite{Dave:2000ar, Vogelsberger:2012ku, Rocha:2012jg, Peter:2012jh, Zavala:2012us}. As DM annihilations are almost forbidden at temperatures of the recombination epoch, this light DM case evades the constraints from the CMB anisotropies, which are sensitive to energy injection into the intergalactic medium \cite{Slatyer:2015jla, Slatyer:2015kla}.

     In the present work, we focus on the parameter region where the rate of the hidden sector interacting with the SM drops below the Hubble expansion at $T\sim m_X$ so that the hidden sector evolves with a different temperature from the SM plasma.  The scenario has the following properties: 
\begin{itemize}
\item  Interactions among the dark sector particles guarantee $T_X=T_S$ before freeze-out.
\item Below $T\sim m_X$, the 3-to-2 cannibal annihilations are still active. The hidden sector temperature may be hotter than the SM bath as the number-changing interactions convert the mass of nonrelativistic hidden sector particles into their kinetic energy. Meanwhile, the hidden sector may cool down efficiently due to net entropy injection from the hidden sector to the SM bath, resulting from the $S$ decaying out-of-equilibrium into the SM bath. 
\item During the cannibalization epoch, the hidden sector densities, following their temperature, are Boltzmann depleted and maintain chemical equilibrium (with zero chemical potential).
\item When cannibalization ends, the forbidden channel described by the DM annihilation to mediators is still active and guides the evolution of hidden sector densities until freeze-out;
meanwhile, 2-to-2 interactions keep hidden sector particles in temperature equilibrium but with a nonzero chemical potential.
\end{itemize}

This paper is organized as follows. In Sec.~\ref{sec:boltz}, we introduce the simplest renormalizable vector dark matter model suitably used in the present study for the forbidden dark matter under a cannibally co-decaying phase. For the condition that the hidden sector is in chemical and kinetic equilibrium with the SM bath before freeze-out,
 we will show that only a much larger mass ratio $m_S/m_X \in (1.5,1.65)$ can simultaneously give the correct relic density and have a proper size of DM self-interactions to explain the small scale problems. However, the corresponding parameter space, related to the mixing angle between the SM-like Higgs and hidden scalar, is ruled by fixed target and flavor experiments.
In Sec.~\ref{sec:cannibal-codecay}, we study a scenario in which the hidden sector is thermally (chemically and kinetically) decoupled from the bath earlier at about $T\sim m_X$.
We develop a formalism relevant to the evolutions of hidden particles. We revise the evolution scale, the bath temperature, which may be affected by a non-thermal equilibrium process due to the hidden scalar's out-of-equilibrium decay, resulting in an increase in total comoving entropy.
We finally obtain the coupled Boltzmann equations for the number densities and temperatures and show how these quantities evolve when the hidden sector undergoes a cannibally co-decaying phase. 
The experimental constraints are presented in Sec.~\ref{sec:phenomenology}, where we show that this scenario can offer a parameter space not only having an ample enough coupling strength between $X$ and $S$ to account for the small scale problems as well as the correct DM relic but also evading the current constraints. 
Experiments can further probe this scenario. The conclusions are contained in Sec.~\ref{sec:conclusions}.

\vskip2cm

\section{ Boltzmann equations and thermal freeze-out}\label{sec:boltz}

To feature the thermal mechanism of the hidden sector, we take the simplest renormalizable vector dark matter model~\cite{Baek:2012se, Yang:2019bvg} as an example to illustrate the pictures. Here, the dark sector contains only a complex scalar $\Phi_S$ and an abelian $U(1)_X$ gauge vector boson
$X_\mu$ (denoted as $X$ throughout this paper).
The covariant derivative of the dark sector is given by $ D_\mu \Phi_S \equiv (\partial_\mu + i g_{\rm dm} Q_{X} X_\mu )\Phi_S$, where $Q_{X}$ is the $U_X(1)$ charge value of $\Phi_S$. The scalar potentials of the Lagrangian are given by
\begin{align}
 V   =   \mu_{H}^2 |\Phi_H|^2 + \mu_{S}^2 |\Phi_S|^2  + \frac{\lambda_H}{2} (\Phi_H^\dagger \Phi_H)^2 
  + \frac{\lambda_S}{2} (\Phi_S^\dagger \Phi_S)^2
  +  \lambda_{HS} (\Phi_H^\dagger \Phi_H) (\Phi_S^\dagger \Phi_S)  \;,
\label{eq:lagrangian}
\end{align}
where $\Phi_H =(H^+, H^0)^{\rm T}$ is the SM Higgs doublet. After spontaneous symmetry breaking (SSB),  
the two Higgs fields can be parametrized in terms of their vacuum expectation values (VEVs), $v_H=246$~GeV and $v_S$, and the real neutral components, $\phi_h$ and $\phi_s$, in the unitary gauge as
$\Phi_H =\frac{1}{\sqrt{2}} (0, v_H + \phi_h )^{\rm T} $ and $\Phi_S=\frac{1}{\sqrt{2}} (v_S + \phi_s)$, and the dark matter gets its mass with $m_X=g_{\rm dm} Q_X v_S$.
The two real scalars are related to the mass eigenstates,
$h  = \cos\alpha \, \phi_h  +  \sin\alpha \, \phi_s$,
$S =- \sin\alpha \, \phi_h  + \cos\alpha \, \phi_s $. We take $h$ as the SM-like Higgs with mass $m_h=125.18$~GeV \cite{pdg2022} and use the dark charge $Q_X=1$ in the analysis. 

Note that after SSB, the $Z_2$ symmetry, $X_\mu \to -X_\mu$ and $\Phi_S\to  \Phi_S^*$, can stabilize the DM persists in the Lagrangian of this simplified model. 
We remark here. The present model is a low-energy effective theory. The Lagrangian could exist some nonrenormalizable higher dimensional operators that, resulting from some other UV complete theory\footnote{For instance, the $Z_2$ symmetry of the DM can be broken by another $U(1)'$ symmetry which is spontaneously broken by a new scalar making the new $U(1)'$ gauge boson with a mass $\sim \Lambda$ \cite{Baek:2014goa}. Meanwhile, the hidden scalar $\Phi_S$ together with SM particles, which can be fermions or the SM Higgs, is charged under $U(1)'$.} beyond a much higher scale $\Lambda \gg v_H, v_s$, break the $Z_2$ symmetry and are thus suppressed by $\Lambda$ at the low energy scale. The existence of the nonrenormalizable higher dimensional operators might lead to the vector DM decay into the SM particles, which could be observed/constrained in the indirect searches \cite{Baek:2014goa, Arina:2009uq, Gustafsson:2013gca}.

In this paper, we consider $m_X<m_S$. Thus, the two-body annihilation $XX\to SS$, forbidden at zero temperature, may occur at finite temperatures, and its thermally averaged rate depends on the mass ratio $r=m_S/m_X$ (for comparison, we will briefly discuss the case of
DM in the form of a Majorana fermion in Appendix~\ref{app:majorana}). The coupled Boltzmann equations describe the evolutions of number densities of $X$ and $S$,
\begin{align}
\frac{dn_X}{dt}  & + 3 H n_X =
  \{2 \leftrightarrow 2 \}_{SX}      
  - \{2 \leftrightarrow 2 \}_{\rm SM}^X + \{3 \leftrightarrow 2 \}_X  , 
   \label{eq:boltz-11} \\
\frac{dn_S}{dt}  & + 3 H n_S =  
-  \bigg(  \langle\Gamma_{S}\rangle_{T_i} n_S -   \langle\Gamma_{S}\rangle_T n_S^{\text{eq}} (T)  \bigg)   -  \{2 \leftrightarrow 2 \}_{SX}    - \{2 \leftrightarrow 2 \}_{\rm SM}^S \,
     + \{3 \leftrightarrow 2 \}_S    \,, 
      \label{eq:boltz-21}
\end{align}
where  
\begin{align}
\{2 \leftrightarrow 2 \}_{SX}  \equiv &
 \langle \sigma v \rangle_{SS\to XX} (T_i)
\bigg( n_S^2 -  \frac{  (n_S^{\text{eq}} (T_i) )^2 }{(n_X^{\text{eq}}(T_i))^2}  n_X^2 \bigg) , \nonumber\\
\{2 \leftrightarrow 2 \}_{\rm SM}^X
 \equiv &
   \langle \sigma v \rangle_{XX\to {\rm SM}\, {\rm SM}} (T_i) \,  n_X^2 
 -   \langle \sigma v \rangle_{XX\to {\rm SM}\, {\rm SM}} (T)  \, (n_X^{\text{eq}} (T) )^2  \,, \nonumber
\end{align}
$\{2 \leftrightarrow 2 \}_{\rm SM}^S$ is the same as $\{2 \leftrightarrow 2 \}_{\rm SM}^X$ but with $X$ replaced by $S$,
$n_{X,S}^{\rm eq}(T_i)$ are the equilibrium densities at temperature $T_i$, and $\langle\Gamma_S\rangle$ is the thermally averaged width of $S$, $\{3 \leftrightarrow 2 \}_{X,S}$  denote terms involving $3 \leftrightarrow 2$ number-changing interactions (see Eqs.~(\ref{eq:boltz-YX}) and (\ref{eq:boltz-YS}) and  Refs.~\cite{Yang:2019bvg,Yang:2022zlh} for a detailed expression). 

\begin{figure*}[t!]
\begin{center}
\includegraphics[width=0.68\textwidth]{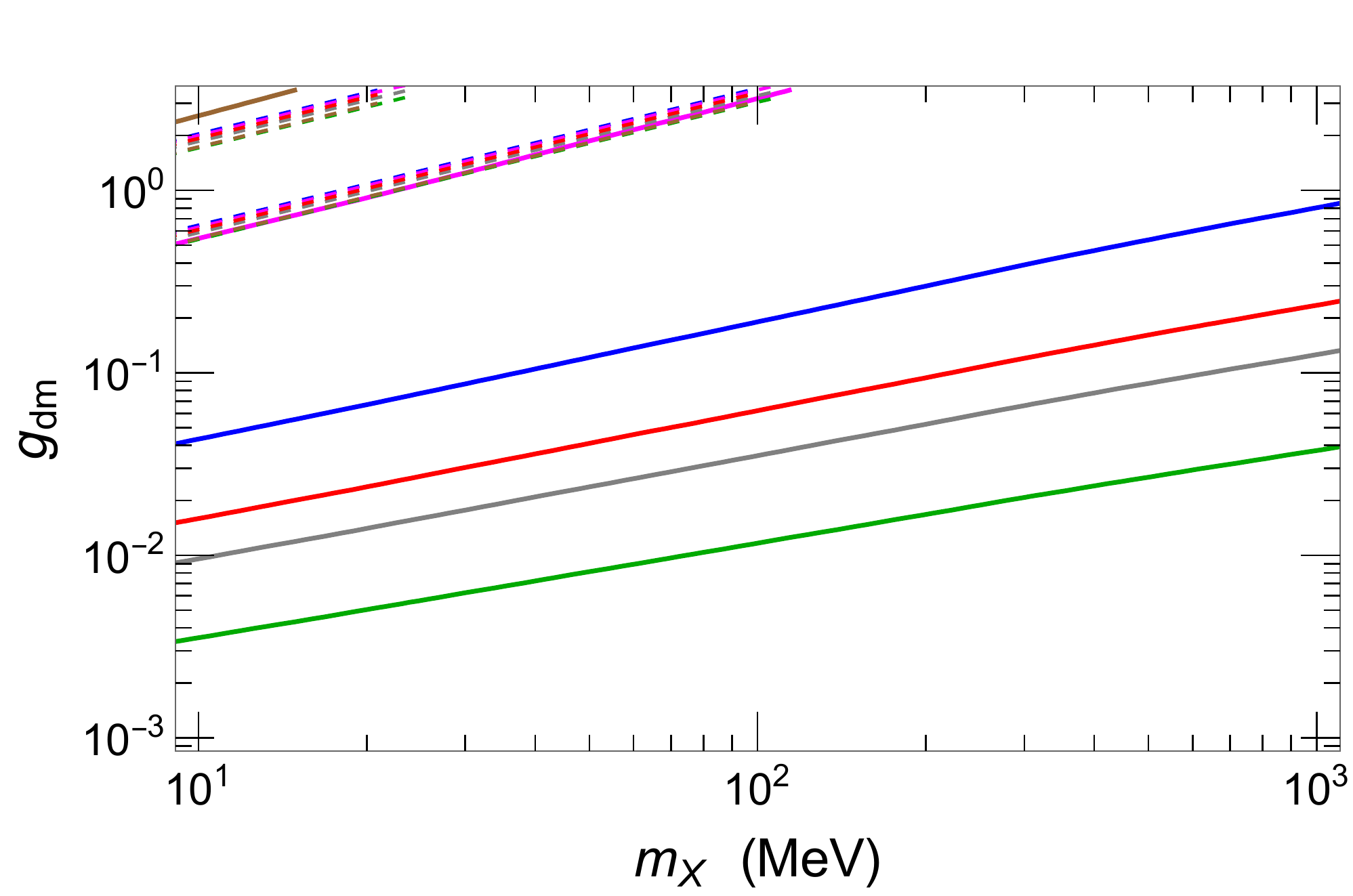}
\caption{$g_{\rm dm}$ as a function of $m_X$. (i) The solid lines, corresponding to different $r$, can produce the correct relic density for the case with a sufficiently sizable $\alpha$ that can maintain the thermal equilibrium between the hidden sector and SM during the freeze-out process. (ii)  The regions bounded respectively by two dashed lines with the same color approximate the current self-interactions $\sigma_{\rm SI}/m_{X} \in (0.1,10)~{\rm cm}^2/g$ and are highly insensitive to the value of $r$.
 The green, gray, red, blue, magenta, and brown lines correspond to $r (=m_S/m_X)=1.01, 1.1, 1.15, 1.25, 1.5$, and 1.65, respectively. The unitarity bound $g_{\rm dm}< \sqrt{4\pi}$ is imposed. Note that only the case with  $r\in (1.5,1.65)$ can simultaneously account for the correct relic density and the requirement of the small-scale problems.}
\label{fig:model-mx-gdm}
\end{center}
\end{figure*}

The decoupling temperature of the SM bath and the hidden sector is related to the coupling strength, relevant to $\alpha$, between these two sectors. For a suitably large value of $\alpha$, the hidden scalar $S$ can be treated as a part of the {\it effective} bath (SM plus $S$) because it keeps thermal equilibrium with the SM particles via 
$\text{SM}~S \leftrightarrow \text{SM}~S$ \footnote{ We have $SS \to \text{SM SM} \sim XX \to  \text{SM SM}$ and $\text{SM}~S \to \text{SM}~S \sim \text{SM}~X \to \text{SM}~X $.
The rate $\text{SM SM}\to SS$ is less than $\text{SM}~S \to \text{SM}~S$ because the former is $s$-channel dominant and suppressed by initial $S$ number density, while the latter is $t$-channel dominant and its initial SM particles could be relativistic for $T<m_S$.}
 and $\text{SM SM}\leftrightarrow S$, together with cannibalization\footnote{The cannibalization and inverse $S$ decay can result in the zero chemical potential of $S$ during freeze-out. }.
Thus, for this condition, the DM relic density, independent of $\alpha$, is set by thermally averaged forbidden $XX\to SS$ annihilation cross section\footnote{We consider the parameter region, where the $XX \to \text{SM SM}$ cross section is much less than the typical cross section with $\sim (3.8-5.0)\times 10^{-9}$~GeV$^{-2}$ in the WIMP scenario for 7~MeV$\lesssim m_X \lesssim 100$~MeV. For comparison, taking $m_X=100$~MeV, $r=1.25$ and $g_{\rm dm }=0.2$, we have $\langle \sigma v \rangle_{XX\to {\rm SM}\, {\rm SM}}=3.1 (g_{\rm dm}/0.2)^2 \sin^2\alpha \times 10^{-11}$~GeV$^{-2}$ at $T\to 0$, which cannot be dominant during DM freeze-out even using $\sin\alpha=1$. Usually, the $s$-channel $XX \to \text{SM SM}$ is dominant during freeze-out when the resonant region, $m_S\sim 2m_X$, is considered. However, this is outside the parameter region that we are interested in. See also Fig.~\ref{fig:relrate} and related discussions for the present study.}, which is specially controlled by the values of the coupling $g_{\rm dm}$ and the mass ratio $r  (=m_S/m_X$).
 In Fig.~\ref{fig:model-mx-gdm}, denoted as the solid lines, we show the coupling strength $g_{\rm dm }$ that accounts for the observed relic density. For a case with a larger $r$, a more prominent $g_{\rm dm }$ is necessary.
For instance, at $m_X=100 $~MeV, $g_{\rm dm }$ is about $0.01$ for $r=1.01$ but becomes $0.2$ for $r=1.25$. 


On the other hand, the small-scale structures favor the DM self-interactions around $\sigma_{\rm SI}/m_{X} = 0.1 -10~{\rm cm}^2/g$.  The cross-section for the self-interacting DM scattering $XX\to XX$ through the scalar mediator exchange via $s$-, $u$-, and $t$-channels can be obtained in the zero-velocity limit of the incoming particles in the present model,  given by (with $c_\alpha\equiv \cos\alpha$)
\begin{align}
\sigma_{\rm SI} = 
\frac{ c_\alpha^4 \,  g_{\rm dm}^4 m_X^2 ( 32 m_X^4  - 24 m_X^2 m_S^2 + 7m_S^4 +2m_S^2 \Gamma_S^2)}{96 m_S^4 \Big( (4m_X^2 -m_S^2)^2 + m_S^2\Gamma_S^2 \Big)\pi}\,.
\end{align}
 In Fig.~\ref{fig:model-mx-gdm}, we use several different values of $r$ as inputs and then show allowed regions  $\sigma_{\rm SI}/m_{X} \in (0.1,10)~{\rm cm}^2/g$ separately bounded by two dashed lines with the same color.  These results with $g_{\rm dm}$ of order one are highly insensitive to the mass ratio $r$.  Note that for the case with a sufficiently sizable $\alpha$ maintaining the thermal  (chemical) equilibrium between the hidden sector and SM during the freeze-out process,
 only a much larger mass ratio $r\in (1.5,1.65)$ can simultaneously account for the correct relic density and the requirement of the small-scale problems.
 Nevertheless, for $r\in (1.5,1.65)$, we need $\alpha>10^{-3}$ for $m_X>10$~MeV to keep thermal equilibrium between the hidden sector and SM during the freeze-out process. 
 The corresponding parameter region is excluded by E949 \cite{BNL-E949:2009dza}, PS191 \cite{Bernardi:1985ny, Bernardi:1986hs, Bernardi:1987ek, Gorbunov:2021ccu}, and CHARM \cite{CHARM:1985anb} measurements.  The result is shown in  
Fig.~\ref{fig:model-xs-mx-gdm} and the related discussion is given in Appendix~\ref{thermal_freezeout}.

\section{Cannibally co-decaying phase}\label{sec:cannibal-codecay}

As shown above, if $g_{\rm dm} \propto {\cal O}(1)$ can account for the small-scale phenomenon, and if the hidden sector and SM are in good thermal equilibrium during freeze-out,  a much larger splitting ratio $r\in (1.5,1.65)$ is required to have a good fit to the observed relic abundance. However, under this condition that the thermal equilibrium is maintained during freeze-out, the corresponding value of $\alpha$, which should be larger than $10^{-3}$ at least, is excluded by experiments (see Fig.~\ref{fig:model-xs-mx-gdm}).

Fortunately, the scenario of a smaller $\alpha$, resulting in the nonrelativistic hidden sector being thermally decoupled from the SM earlier, not only allows a significant parameter space (with $g_{\rm dm} \propto {\cal O}(1)$) of explaining the correct DM relic density as well as the small scale structures but also can evade the current experimental constraints.
Dynamically, when the hidden sector is decoupled from the SM earlier at $T\sim m_X$, the hidden sector enters a cannibal phase and evolves with an independent temperature, which is higher than the bath due to receiving heat through $3\to 2$ processes. For this scenario, we have to increase the coupling strength ($g_{\rm dm}$) between $X$ and $S$ so that the DM density can be sufficiently depleted with the out-of-equilibrium decay of $S$ to have a relic abundance consistent with the observation.
We will give a detailed analysis in the following, starting with a study of the thermal evolution of the hidden sector.

\begin{figure}[t]
  \hspace{-0.3cm}
  \includegraphics[width=10.0cm]{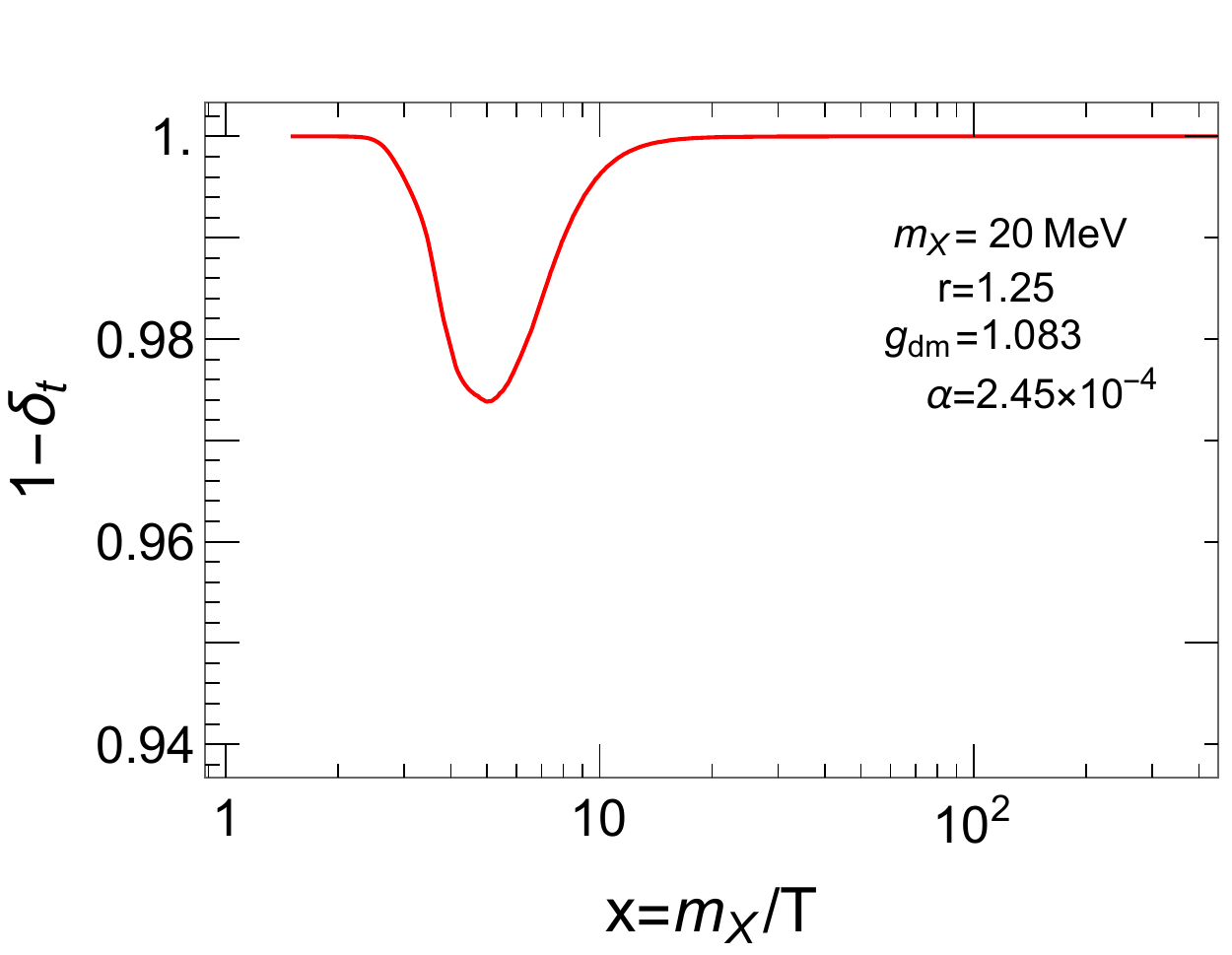}
  \caption{An additional factor $1-\delta_t$ is needed to translate the time scale to the temperature parameter $x$. This is due to the increase of the total comoving entropy resulting from the out-of-equilibrium decay of the scalar $S$.}
  \label{sup-fig}
  \vspace{0cm}
\end{figure}

Before freeze-out, we consider the dark sector particles in thermal equilibrium, i.e., $T_X=T_S$.
As for the condition that $S$ decays out of equilibrium, which results in net entropy injection from the hidden sector to the bath, the SM bath is reheated; as a result, following the second law of thermodynamics, the total comoving entropy $S_t$ is also increased. 
Therefore, we can give the change of the total comoving entropy in the general form,
\begin{align}
  d S_t &= d(s a^3)= \frac{dQ}{T} -  \frac{dQ}{T_X}\,.
\end{align}
The net heat transferred from the hidden scalar to the SM bath is described by
\begin{align}
dQ = -d (a^3 \rho_S) - p_S\, d(a^3)
       &= - a^3 dt  \left[\dot{\rho}_S + 3 \frac{\dot{a}}{a} (\rho_S+p_S) \right]\nonumber \\
      &=   a^3 \, dt\,   \Gamma_S m_S  \left(   n_S - \,  n_S^{\rm eq}(T)   \right) \,,
\end{align}
where $\rho_S$ and $p_S$ are the energy density and pressure of $S$, respectively, $n_S^{\rm eq}(T)$ is the equilibrium density of $S$ at the bath temperature $T$,  and  $a$ is the cosmic scale factor. Here, the total entropy density of the Universe is
\begin{align}
 s= \frac{2\pi^2}{45} h_{\rm eff}(T) T^3 \,,
\end{align}  
where the effective number of the total degrees of freedom is given by
 \begin{align}
 h_{\rm eff}(T) =h_{\rm eff}^{\rm SM}(T) + \left( h_{\rm eff}^X(T_X) + h_{\rm eff}^S (T_X) \right) \left( \frac{T_X}{T} \right)^3 
 \end{align}
with
 \begin{align}
  h_{\rm eff}^i(T_X) = \frac{45 g_i}{4\pi^4} 
  \Big( \frac{m_i}{T_X} \Big)^4 \int_1^\infty \frac{y (y^2-1)^{1/2}}{e^{m_i y/T_X} -1}  \frac{ 4y^2 -1}{3y} dy \,,
  \label{dof-hidden}
 \end{align}
 and $i\equiv X$ or $S$ for the DM or hidden scalar.
  Including the effect that the total comoving entropy could be changed due to the out-of-equilibrium decay of the hidden scalar and considering that the temperature decoupling of $S$ and $X$ occurs until after freeze-out,
we can then deduce the time derivative in terms of the bath temperature,
\begin{align}
\frac{d}{dt}
    =\frac{ds}{dt} \frac{dT}{ds}  \frac{d}{dT}
  &  = \left[ \frac{1}{3s} \Gamma_S\,  m_S \left( n_S -  n_S^{\rm eq}(T) \right) (1-y^{-1})
          -H T \right] \frac{h_{\rm eff}(T)}{\tilde{h}_{\rm eff}(T)}  \frac{d}{dT}\nonumber\\
   &= \left( 1 - \delta_t  \right)   xH
   \frac{h_{\rm eff}(T)}{\tilde{h}_{\rm eff}(T)}  \frac{d}{dx}\,,
\end{align}
where $x = m_X / T$ is the dimensionless temperature variable, $H  = [(8\pi /3) G_N \rho]^{1/2}$ is the Hubble parameter with $\rho$ the total energy density of the Universe, $\tilde{h}_{\rm eff}  \equiv h_{\rm eff} [1+(1/3) (d\ln h_{\rm eff} / d\ln T)]$, and
\begin{align}
\delta_t   \equiv   \frac{1}{3 H } x \frac{m_S}{m_X} 
      \Gamma_S    \left( Y_S - Y^{\rm eq}_S(T_X) \right)  (1-y^{-1})  \,,
\end{align}
with $y\equiv T_X/T$. Here and through this paper, we introduce dimensionless quantities,
$Y_X \equiv n_X /s $ and $Y_S \equiv n_S /s$ for yields, and 
$y= T_X /T$ for the temperature ratio as functions of the dimensionless variable $x$.  In Fig.~\ref{sup-fig}, we show the relative factor $1-\delta_t$, which results from the increase of the total comoving entropy.

As for the present case, the hidden sector is decoupled from the SM bath at $T \sim m_X$ and then evolves with an independent temperature. Thus, to obtain the detailed thermal evolutions of the hidden particles, we need to solve the coupled equations for $Y_X$, $Y_S$, and their temperature dependence $y=T_X/T$.
The Boltzmann equations of yields are given by
\begin{align}
\frac{d Y_X}{dx} =
   & (1-\delta_t)^{-1}
  \frac{\tilde{h}_{\rm eff}(T)}{h_{\rm eff}(T)}   \frac{1}{xH}  
   \Bigg[
      3 H \delta_t Y_X
         + s \langle \sigma v\rangle_{SS\to XX} (T_X)  \bigg( Y_S^2
         -   \frac{(Y_S^{\text{eq}} (T_X))^2 }{ (Y_X^{\text{eq}} (T_X)^2}   Y_X^2  \bigg)  \nonumber\\
     & \
        + \frac{s^2}{3} \langle \sigma v^2 \rangle_{SSS\to XX}
                 \bigg( Y_S^3 - \frac{(Y_S^{\text{eq}} (T_X))^3 }{(Y_X^{\text{eq}} (T_X))^2}  Y_X^2  \bigg)
           - s^2 \langle \sigma v^2 \rangle_{XXS\to SS}
                  \bigg( Y_X^2 Y_S -    \frac{(Y_X^{\text{eq}} (T_X))^2}{Y_S^{\text{eq}} (T_X)}  Y_S^2   \bigg)
            \nonumber\\
     &  \
         - \frac{s^2}{3} \langle \sigma v^2 \rangle_{XXX\to XS}
          \bigg( Y_X^3 -\frac{  (Y_X^{\text{eq}}(T_X))^2 }{ Y_S^{\text{eq}}(T_X) }   Y_X   Y_S \bigg)
             \nonumber\\
     & \
         -   s \Big(  \langle \sigma v \rangle_{XX\to \sum_{ij} {\rm SM}_i {\rm SM}_j}  (T_X) Y_X^2
            - \langle \sigma v \rangle_{XX\to \sum_{ij} {\rm SM}_i {\rm SM}_j} (T) \, ( Y_X^{\text{eq}} (T) )^2 \Big)
        \Bigg] \,,
  \label{eq:boltz-YX}
 \end{align}
\begin{align}
\frac{d Y_S}{dx} =
 & (1-\delta_t)^{-1}
   \frac{\tilde{h}_{\rm eff}(T)}{h_{\rm eff}(T)}   \frac{1}{xH}  \nonumber\\
 &\times \Bigg[
   3 H \delta_t Y_S
      -s \langle \sigma v\rangle_{SS\to XX} (T_X)  \bigg(    Y_S^2
      -   \frac{(Y_S^{\text{eq}} (T_X))^2 }{ (Y_X^{\text{eq}} (T_X)^2}   Y_X^2  \bigg)  \nonumber\\
 &  -  \Gamma_{S}    \bigg( \frac{K_1(m_S/T_X)}{K_2( m_S/T_X)} Y_S
         - \frac{K_1(m_S/T)}{K_2( m_S/T)} Y_S^{\text{eq}}(T) \bigg)
   \nonumber\\
 &  -    s \Big(  \langle \sigma v \rangle_{SS\to \sum_{ij} {\rm SM}_i {\rm SM}_j}(T_X)  Y_S^2
         - \langle \sigma v \rangle_{SS\to \sum_{ij} {\rm SM}_i {\rm SM}_j} (T) \, ( Y_S^{\text{eq}} (T) )^2 \Big)
   \nonumber\\
 &
         +
              \frac{s^2}{6}  \langle \sigma v^2 \rangle_{XXX\to XS}
                      \bigg( Y_X^3 -  Y_X  Y_S    \frac{ (Y_X^{\text{eq}}(T_X))^2 }{ Y_S^{\text{eq}}(T_X)}   \bigg)
             - \frac{s^2}{6}    \langle \sigma v^2 \rangle_{SSS\to SS}
              \Big( Y_S^3 - Y_S^2  Y_S^{\text{eq}} (T_X) \Big)
     \nonumber\\
 &
      +    \frac{s^2}{2}  \langle \sigma v^2 \rangle_{XXS\to SS}
                   \bigg( Y_X^2 Y_S -   \frac{(Y_X^{\text{eq}} (T_X))^2 }{Y_S^{\text{eq}}(T_X)} Y_S^2 \bigg)
  -    \frac{s^2}{2}  \langle \sigma v^2 \rangle_{SSS\to XX}
               \bigg( Y_S^3 - \frac{ (Y_S^{\text{eq}}(T_X))^3}{(Y_X^{\text{eq}}(T_X))^2} Y_X^2 \bigg)
     \nonumber\\
 &   -  \frac{s^2}{2}  \langle \sigma v^2 \rangle_{XSS\to XS}
              \Big( Y_X Y_S^2 - Y_X Y_S  Y_S^{\text{eq}}(T_X)  \Big)
\Bigg]
    \,.  \label{eq:boltz-YS}
 \end{align}

As for the temperature evolution, considering that the hidden species evolve with the same temperature, i.e., $T_X=T_S$, and adopting the definition
\begin{align}
T_X =\frac{g_X}{n_X } \int \frac{d^3 p_X}{(2\pi)^3} \frac{{\bf p}_X^2}{3 E_X}  f_X (T_X) \,,
\label{eq:temp-def}
\end{align}
we can then obtain the Boltzmann moment equation for the hidden temperature,
\begin{align}
     (n_X  + n_S) \frac{d T_X}{dt}
      + 
     \Big( & (2-\delta^X_H) n_X +   (2-\delta^S_H) n_S  \Big)   H  T_X  \nonumber  \\
   = 
     &- \left( \frac{d n_X }{dt} +3 H n_X  \right) T_X
          + g_X \int  \frac{d^3p_X}{(2\pi)^3}  \, C \Big[f_X\cdot \frac{{\bf p}_X^2}{3 E_X} \Big]  \nonumber\\
              & - \left( \frac{d n_S }{dt} +3 H n_S  \right) T_X
          + g_S \int  \frac{d^3p_S}{(2\pi)^3}  \, C \Big[f_S\cdot \frac{{\bf p}_S^2}{3 E_S} \Big]
     \,.
       \label{eq:boltz-t-hidden}
\end{align}
where $f_i$ is the distribution of the species ``{\it i}" in the momentum space, $C [f_i\cdot \frac{{\bf p}_i^2}{3 E_i}]$ is the collision term, and
\begin{align}
\delta_H^i
& \equiv 1- \frac{g_i m_i^2}{n_i \, T_i} \int \frac{d^3 p_i}{(2\pi)^3} \frac{{\bf p}_i^2 }{3 E_i^3}  f_i (T_i) \,,
\label{eq:deltah}
\end{align}
with $i\equiv X, S$. By calculating all relevant collision terms, this temperature equation can be reformulated in the form,
\begin{align}
\frac{dy}{dx} =
& \frac{y}{x}
 -  3\frac{\delta_t}{1-\delta_t}  \frac{\tilde{h}_{\rm eff}(T)}{h_{\rm eff}(T)}
  \frac{y}{x}
 - \frac{y}{Y_X + Y_{S}}
\bigg( \frac{dY_X}{dx} +  \frac{dY_S}{dx}
\bigg)
\nonumber \\
&  +
\frac{(1-\delta_t)^{-1}}{Y_X + Y_{S}}
 \frac{\tilde{h}_{\rm eff}(T)}{h_{\rm eff}(T)}
\Bigg\{
 -  (2 - \delta_{H}^X)
  \Big(\frac{y}{x} +  \frac{\gamma_X}{x H}  (y-1)
  \Big) Y_X
  \nonumber\\
 & \  \
  - \frac{s}{ x H }
     \Big(y\,  Y_X^2   \widetilde{\langle \sigma v \rangle}_{XX\to \sum_{ij} {\rm SM}_i {\rm SM}_j} (T_X)
     -  (Y_X^{\text{eq}} (T) )^2   \widetilde{\langle \sigma v \rangle}_{ XX\to \sum_{ij} {\rm SM}_i {\rm SM}_j } (T)   \Big)
   \nonumber \\
&  \  \
  +  \frac{s^2}{m_X H}   \Bigg[
    \frac{(4m_X^2-m_S^2)(16m_X^2 -m_S^2)}{108 m_X(10m_X^2 -m_S^2)}
     \langle \sigma v^2 \rangle_{XXX\to XS}
    \bigg( Y_X^3- \frac{  Y_X Y_S  \big( Y_X^{\text{eq}} (T_X) \big)^2}{ Y_S^{\text{eq}}(T_X) } \bigg)
    \nonumber\\
& \  \  \  \
  + \frac{ m_S^2 (2m_X+m_S)(2m_X+3m_S)}{4(m_X+2m_S)(2m_X^2 +4m_X m_S + 3m_S^2)}
    \langle \sigma v^2 \rangle_{XSS\to XS}
    \Big( Y_X Y_S^2 - Y_X Y_S Y_S^{\text{eq}}(T_X)  \Big)
 \nonumber\\
& \  \  \  \
  + \frac{(2m_X +3m_S)(3m_S -2m_X)}{54 m_S}
     \langle \sigma v^2 \rangle_{SSS\to XX}
    \bigg( Y_S^3- \frac{  Y_X^2  \big( Y_S^{\text{eq}} (T_X) \big)^3}{ \big( Y_X^{\text{eq}}(T_X)  \big)^2 } \bigg)   \Bigg]
    \nonumber\\
& \  \
 -  (2 - \delta_{H}^S)
   \Big(\frac{y}{x} +  \frac{\gamma_S}{x H}  (y-1)
   \Big) Y_S
   \nonumber\\
&  \  \
 - \frac{s}{ x H }
    \Big(y\,  Y_S^2   \widetilde{\langle \sigma v \rangle}_{SS\to \sum_{ij} {\rm SM}_i {\rm SM}_j} (T_X)
     -  (Y_S^{\text{eq}} (T) )^2   \widetilde{\langle \sigma v \rangle}_{ SS\to \sum_{ij} {\rm SM}_i {\rm SM}_j } (T)   \Big)
   \nonumber \\
&  \  \
    - \frac{ \Gamma_{S}}{x H}
        \bigg(    \frac{K_1(m_S/T_X)}{K_2(m_S/T_X)} Y_S \delta_\Gamma (T_X)  y
    -\frac{K_1(m_S/T)}{K_2(m_S/T)}  Y_S^{\text{eq}} (T)  \delta_\Gamma (T)
         \bigg)
   \nonumber\\
 &  \   \
   +  \frac{s^2}{m_X H}   \Bigg[
     \frac{(4m_X^2-m_S^2)(16m_X^2 -m_S^2)}{108 m_X(8m_X^2+m_S^2)}
     \langle \sigma v^2 \rangle_{XXX\to XS}
     \bigg( Y_X^3- \frac{  Y_X Y_S  \big( Y_X^{\text{eq}} (T_X) \big)^2}{ Y_S^{\text{eq}}(T_X) } \bigg)
    \nonumber\\
&\  \  \  \
 +  \frac{(2m_X+3m_S)(2m_X-m_S)}{6 (2m_X+m_S)}
    \langle \sigma v^2 \rangle_{XXS\to SS}
    \bigg( Y_X^2 Y_S - \frac{Y_S^2 \big( Y_X^{\text{eq}}(T_X) \big)^2}{Y_S^{\text{eq}} (T_X) }  \bigg)
 \nonumber\\
& \  \  \  \
+ \frac{ m_S(2m_X+m_S)(2m_X+3m_S)}{4(m_X+2m_S)(4m_X+5m_S)}
    \langle \sigma v^2 \rangle_{XSS\to XS}
   \Big( Y_X Y_S^2 - Y_X Y_S Y_S^{\text{eq}}(T_X)  \Big)
 \nonumber\\
& \   \  \  \
+ \frac{5}{54}  m_S
 \langle \sigma v^2 \rangle_{SSS\to SS}
   \Big( Y_S^3 - Y_S^2  Y_S^{\text{eq}}(T_X)  \Big)
   \Bigg]
   \Bigg\}
 \,,
\end{align}
Here $\gamma_{X,S}$ parametrize the contributions for elastic scatterings,  $Y_k^{\rm eq}(T_\ell)  \equiv n_k^{\rm eq}(T_\ell) /s (T)$ with the subscript $k\equiv X, S$ and $T_\ell =T_X, T$, and we have used the notations: $\widetilde{\langle \sigma v \rangle}_{SS\to \sum_{ij} {\rm SM}_i {\rm SM}_j} (T_\ell)
\equiv  \langle \sigma v \cdot \frac{{\bf p}_S^2}{3 E_S} \rangle_{S S \to \sum_{ij} \text{SM}_i \text{SM}_j} (T_\ell) /T_\ell$, with the subscript $k\equiv X, S$ and $T_\ell =T_X, T$. The detailed formulas for the relevant cross sections can be found in Refs.~\cite{Yang:2019bvg,Yang:2022zlh}.

\begin{figure*}[t!]
\begin{center}
\vskip-0.35cm
\includegraphics[width=0.47\textwidth]{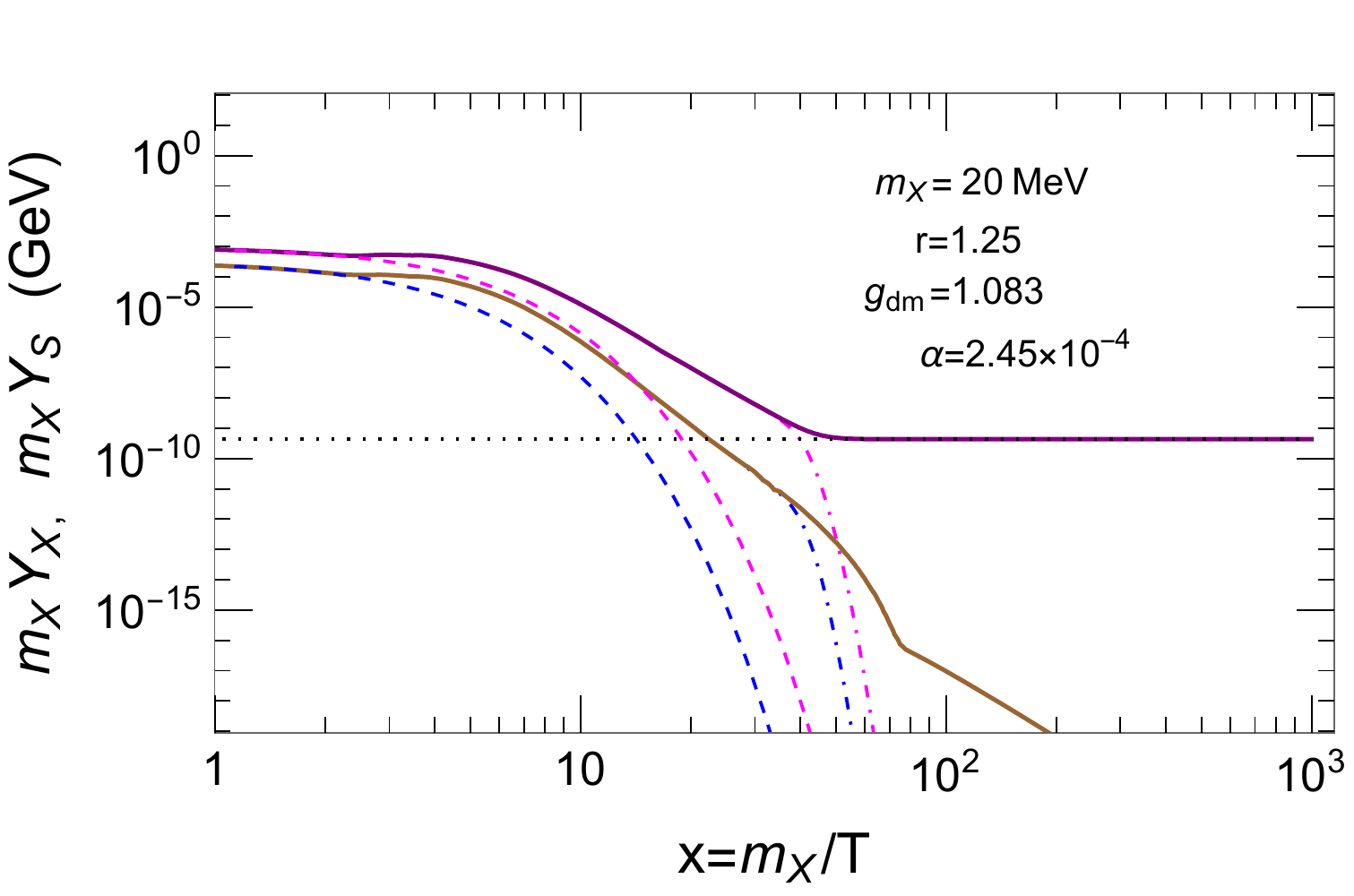}\hskip0.5cm
\includegraphics[width=0.46\textwidth]{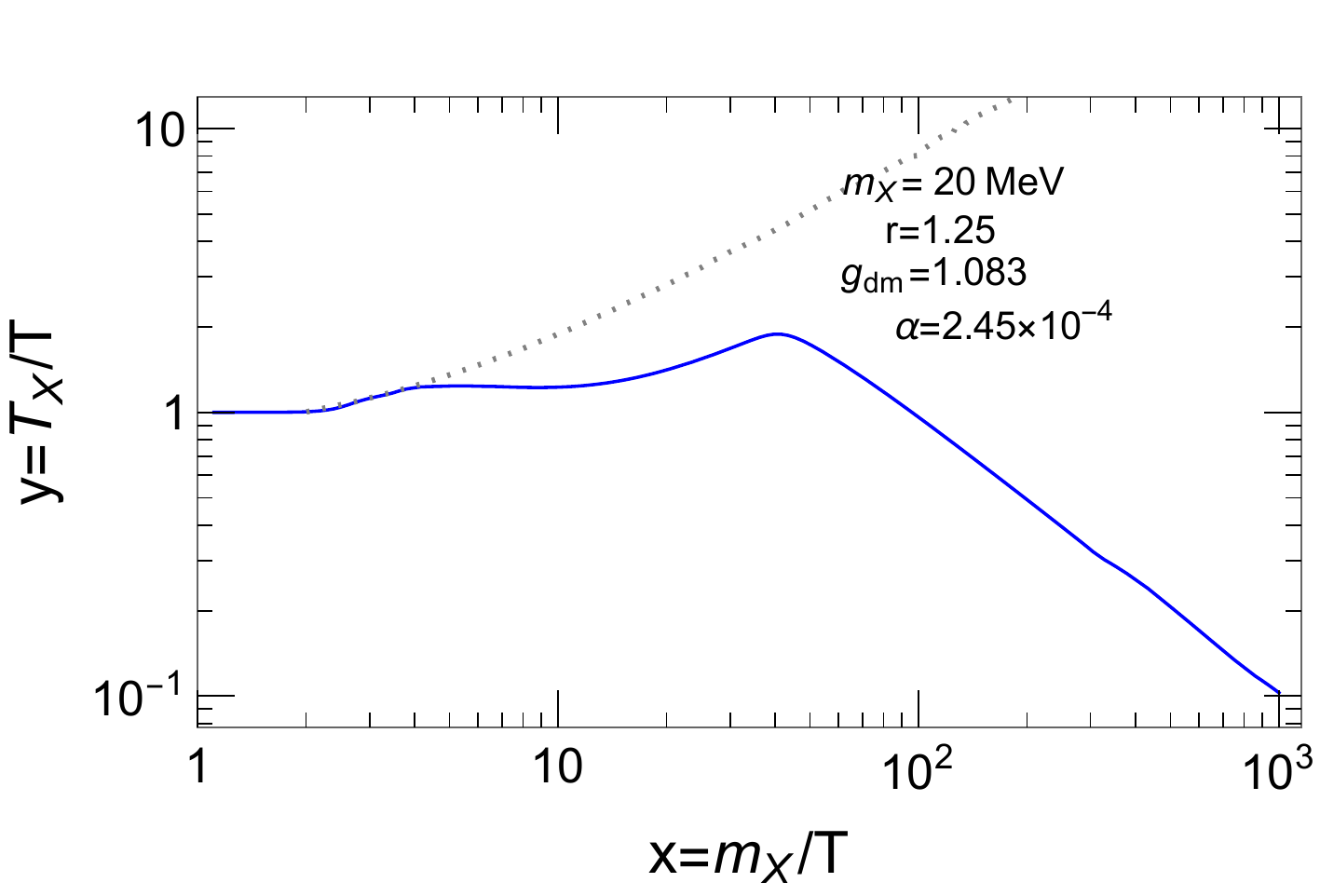}\\
\vskip-0.35cm
\caption{  Left panel: $m_X Y_X$ (solid purple line) and $m_X Y_S$ (solid brown line) as functions of $x$. The magenta and blue dashed lines (or dot-dashed lines) depict the corresponding ones following Boltzmann suppression with $T$ (or their temperature $T_X$). The value of $g_{\rm dm}$ gives $\sigma_{\rm SI}/m_{X} =0.1~{\rm cm}^2/g$.  The horizontal dotted line denotes the correct DM relic abundance, $m_XY_X^\infty =4.37\times 10^{-10}$~GeV.
Right panel:  $y$ vs. $x$ (blue curve). After decoupling, if the entropy is separately conserved in the hidden sector and SM, the temperature ratio instead follows the dashed line. }
\label{fig:relic-boltz1-1}
\end{center}
\end{figure*}

\begin{figure}[t!]
\begin{center}
\hskip-0.3cm\includegraphics[width=0.475\textwidth]{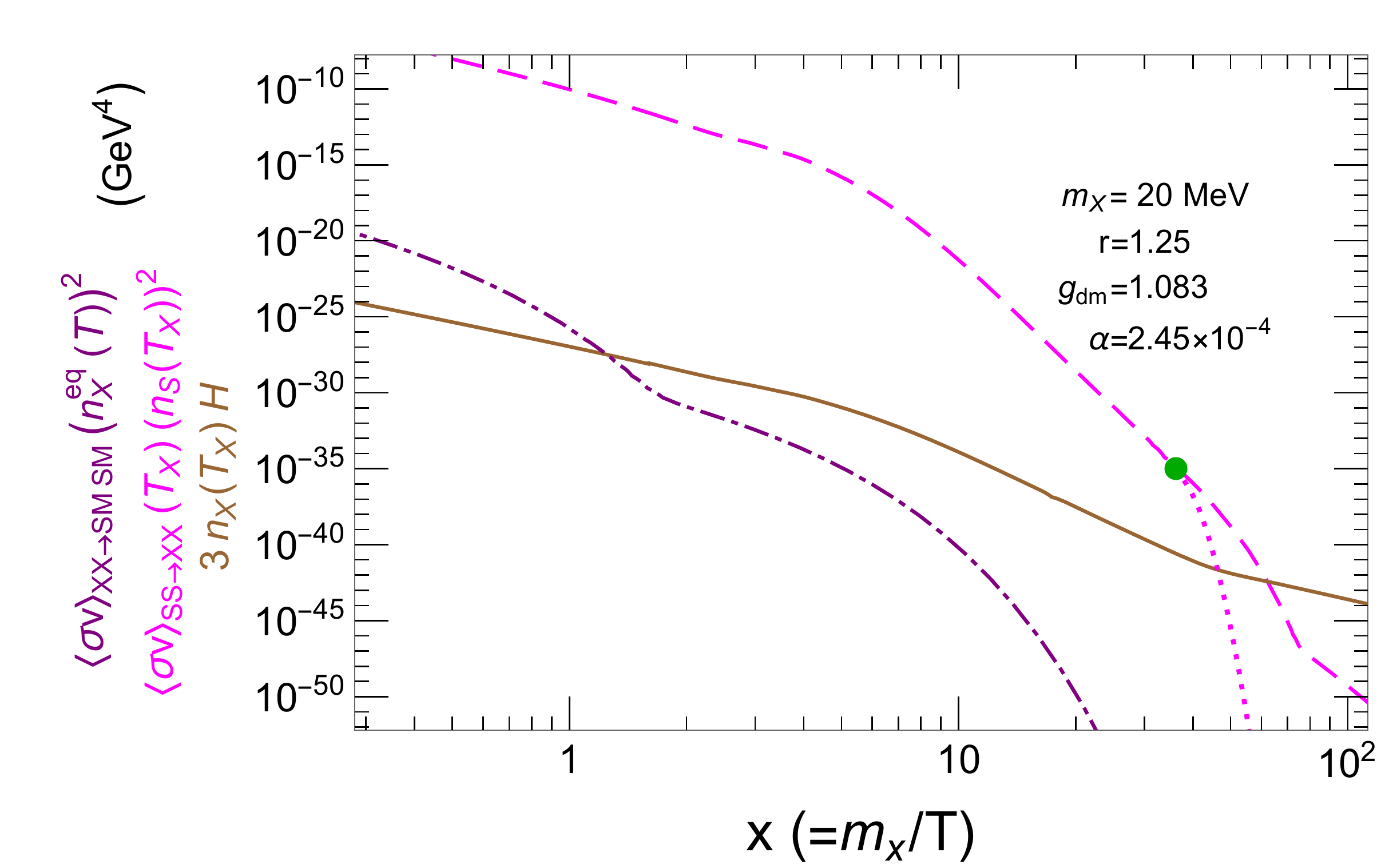}\hskip0.5cm
\includegraphics[width=0.475\textwidth]{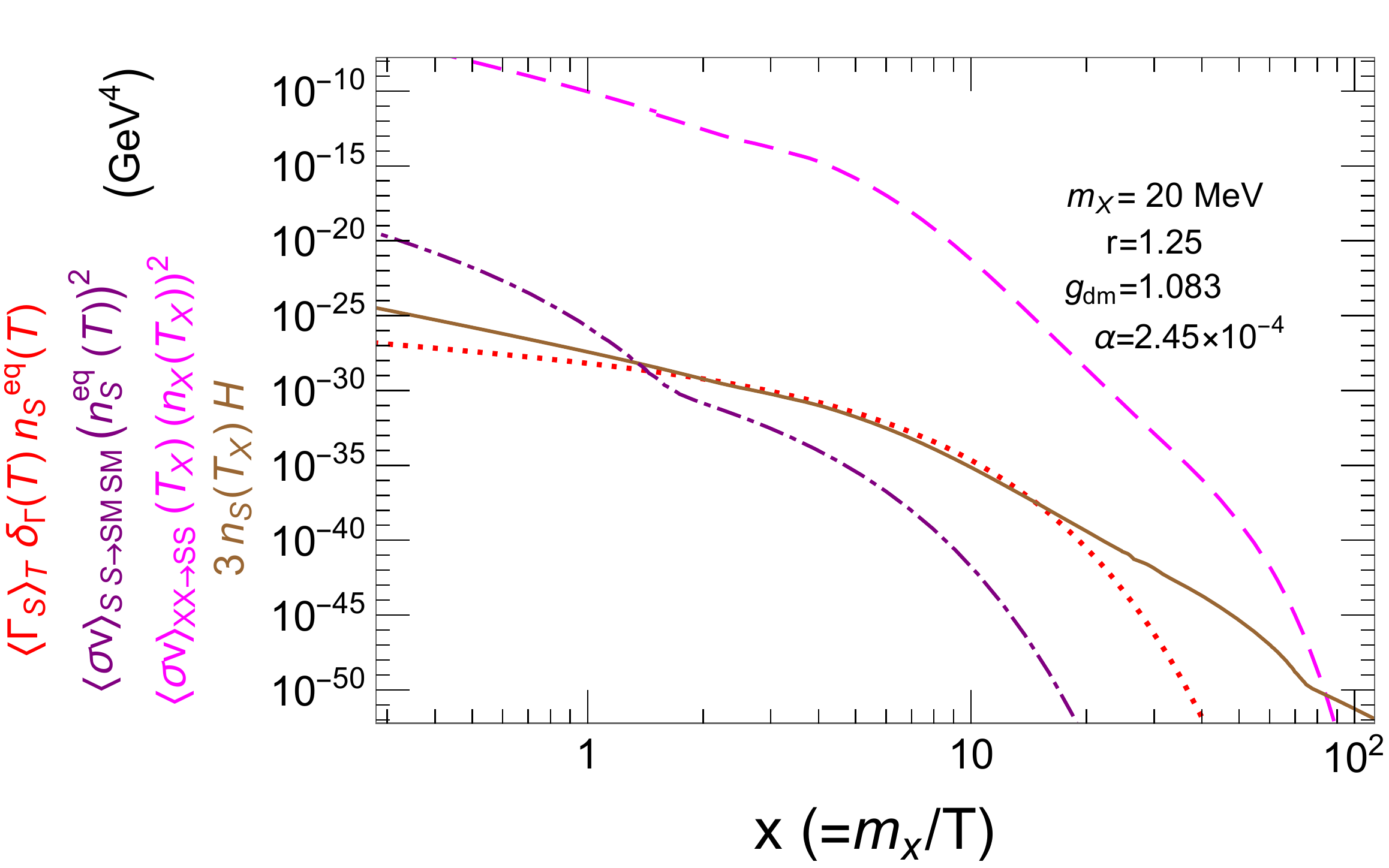} \\
\includegraphics[width=0.6\textwidth]{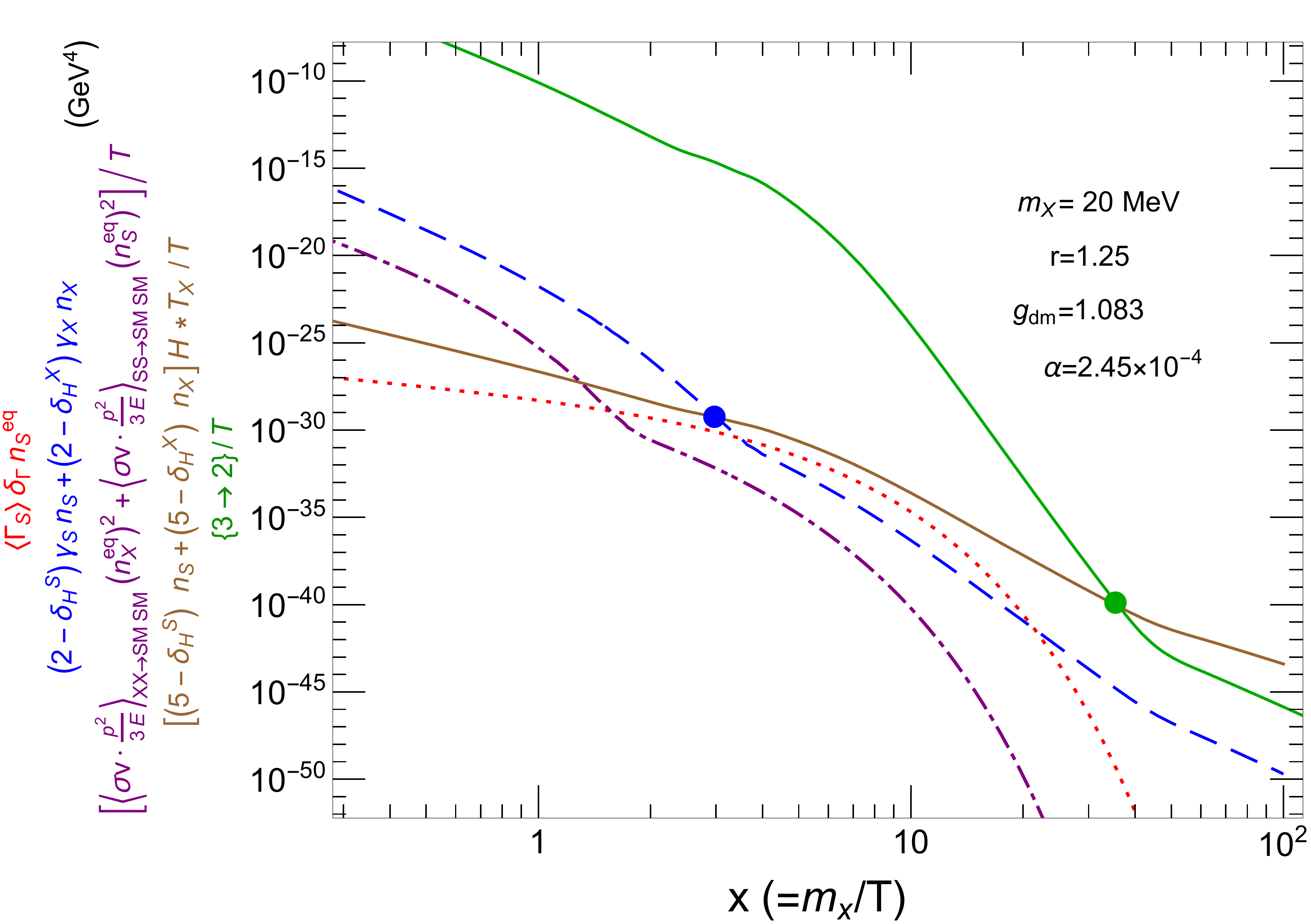} \\
\caption{
Upper left panel: DM number injection rates via $\text{SM~SM}\to XX$ (purple dash-dotted) and $SS \to XX$  (magenta dashed) vs. the dilute rate (brown solid) due to the cosmic expansion.
Upper right panel: $S$ number injection rates via $\text{SM~SM}\to SS$ (purple dash-dotted), $\text{SM~SM}\to S$ (red dot) and $XX \to SS$  (magenta dashed) vs. the dilute rate (brown solid) due to the cosmic expansion.
Lower panel:
Kinetic energy injection rates from the SM bath to hidden sector via annihilations, $\text{SM~SM}\to XX$ and $\text{SM~SM}\to SS$ (purple dash-dotted), elastic scatterings, $\text{SM}~S \to \text{SM}~S$ and $\text{SM}~X \to \text{SM}~X$ (blue dashed), inverse $S$ decays, $\text{SM~SM}\to S$ (red dotted) vs. the cooling rate due to the cosmic expansion (brown solid), together with the heating rate among the hidden particles through $3\to 2$ cannibal processes (green solid), where a factor of $1/T$ has scaled all magnitudes.  
At the temperature drops below that denoted by the blue dot, the elastic scattering rate is approximately less than the cosmic rate. $T_X^c$ corresponds to the green dot; after that, the DM number density then evolves with nonzero chemical potential ($\mu$) instead of following the magenta dotted line with vanishing $\mu=0$. }
\label{fig:relrate}
\end{center}
\end{figure}

As shown in Fig.~\ref{fig:relic-boltz1-1}, imposing $g_{\rm dm}=1.083$, which corresponds to $r=1.25$, $m_X=20$~MeV and $\alpha=2.45\times 10^{-4}$, not only leads to the correct relic density but also indeed results in a sizable $\sigma_{\rm SI}/m_{X} =0.1~{\rm cm}^2/g$. Here $\alpha$ is less than its minimum value\footnote{It is $3.28\times10^{-4}$ for $r=1.25$, $m_X=20$~MeV. See also Fig.~\ref{fig:model-xs-mx-gdm}.} of maintaining the hidden sector and bath in the thermal equilibrium before freeze-out.

In Fig.~\ref{fig:relic-boltz1-1}, we numerically illustrate the typical thermal evolution of the self-interacting forbidden DM under a cannibally co-decaying phase. 
After the (elastic) decoupling temperature, which is $T_{\rm dec} \simeq 8$~MeV corresponding to $x_{\rm dec} \simeq 2.5$ (approximately consistent with the blue dot in Fig.~\ref{fig:relrate}; see also Fig.~\ref{sup-fig}), 
 the active 3-to-2 annihilations result in the hidden sector evolving with a different temperature from the SM bath and meanwhile keep the number densities of the hidden sector particles in chemical equilibrium with zero chemical potential at their temperatures $T_X$, i.e., $n_X=n_X^{\rm eq}(T_X)$ and $n_S=n_S^{\rm eq}(T_X)$.

If there is no net entropy flow between the hidden sector and SM, their entropy ratio is fixed so that the temperature ratio, as the dashed line denoted in the right panel of Fig.~\ref{fig:relic-boltz1-1}, will evolve according to the relation,
\begin{align}
\frac{T_X}{T} =\Big(\frac{s_X+ s_S}{s_{\rm SM}} \Big)^{1/3} 
   \Bigg(    \frac{h_{\rm SM}^{\rm eff}(T) }{ h^{\rm X}_{\rm eff}(T_X) + h^{\rm S}_{\rm eff}(T_X)}  
   \Bigg)^{1/3} \,, \label{eq:hidden-entropy}
\end{align}
where $s_X$ and $s_S$ are the entropy densities of $X$ and $S$, respectively, and their effective numbers of DoF have been given in the form of Eq.~(\ref{dof-hidden}).
In reality, because $S$ decays out-of-equilibrium, resulting in net entropy injection from the hidden sector to the bath, the hidden sector is cooled down, and the SM bath is reheated. 
As such, the number densities $n_{X, S}^{\rm eq}(T_X)$ are further reduced since $T_X$ becomes relatively smaller.
 Moreover, during the cannibalization epoch, because $n_i = n_i^{\rm eq}(T_X) > n_i^{\rm eq}(T)$ with $i\equiv X, S$, the comoving number density decreases with the time, 
 \begin{align}
 \frac{d [(n_X+n_S)a^3] }{dt}   \approx  -   \langle\Gamma_{S}\rangle_{T_X} n_S   \,,
      \label{eq:boltz-decreasing_ni}
\end{align}
which can be obtained from Eqs.~(\ref{eq:boltz-11}) and (\ref{eq:boltz-21}). Therefore, in this scenario, we can expect that a much larger $g_{\rm dm}$ shall be necessary for maintaining a longer period of cannibalization so that the DM can be sufficiently Boltzmann depleted with out-of-equilibrium decay of $S$ to be consistent with the correct relic abundance. For instance, we have $g_{\rm dm}=1.083$ for the case in Fig.~\ref{fig:relic-boltz1-1}, compared to $g_{\rm dm} = 0.07$ with the same $m_X=20$~MeV, $r=1.25$ but under the condition of keeping the hidden sector and the bath in chemical equilibrium before freeze-out (requiring that $\alpha > 3.28\times 10^{-4}$), as shown in Fig.~\ref{fig:model-mx-gdm}.
In Fig.~\ref{fig:relrate}, we sketch the evolutions of the underlying injection rates of number densities and kinetic energy for various reactions compared with the dilute or cooling rate due to the cosmic expansion. The relevant formulas are referred to Eqs.~(\ref{eq:boltz-11}), (\ref{eq:boltz-21}) and (\ref{eq:boltz-t-hidden}). Some related results can be found in Eqs.~(C5) and (C6) of Ref.~\cite{Yang:2022zlh}.
Here, the kinetic energy dilute rate, $[(5 -\delta_H^S) n_S + (5 -\delta_H^X) n_X ] H T_X$, has included the effect due to the change of comoving number densities.
The annihilation rates, $\text{SM SM}\to XX$ and $\text{SM SM}\to SS$, fall below the cosmic expansion rate at $x\sim 1-2$. We have $\langle \sigma v \rangle_{XX\to {\rm SM}\, {\rm SM}}=1.35 (g_{\rm dm}/1.803)^2 (\sin\alpha/0.000245)^2 \times 10^{-17}$~GeV$^{-2}$ at $T\to 0$, which is much less than the case that the relic abundance is accounted for by $\langle \sigma v \rangle_{XX\to {\rm SM}\, {\rm SM}}$ with the value of $\sim 4.4 \times 10^{-9}$~GeV$^{-2}$. $XX\to {\rm SM}\, {\rm SM}$ is dominated by the $s$-channel via a virtual $S$ or $h$; the relevant formulas can be found in Appendix~A2 of Ref.~\cite{Yang:2022zlh}.  For $4~{\rm MeV}\leq m_X \leq m_\mu$. the main annihilation mode of DM is $e^+ e^-$\,\footnote{$\sigma v_{XX\to ff} \propto  [\Gamma_S ]_{m_S \to \sqrt{s}} ^{S\to ff}$, where the invariant mass replaces $m_S$ in the $S\to ff$ partial width. }. In Fig.~\ref{fig:br-life}, we show the branching ratios and lifetime of $S$ in the sub-GeV range.

At the end of this section, we offer a different way to understand the DM evolution under the cannibally co-decaying phase.
For this scenario, the {\it nonrelativistic} hidden sector particles' temperatures are heated by cannibalization, resulting in larger thermal equilibrium densities which evolve with a higher $T_X$ instead of $T$ (see the left panel of Fig.~\ref{fig:relic-boltz1-1}).
From Eq.~(\ref{eq:boltz-YX}), and using the new variable $x_X\equiv m_X/T_X$, and
\begin{align}
dx =d x_X \, y  \left(1- \frac{d \log{y}}{d\log{x}} \right)^{-1} \,,
\end{align}
the evolution of DM after the cannibal decoupling (for which the $3\to2$ rate falls below the Hubble expansion rate), following its temperature, can be approximated by
\begin{align}
\frac{d Y_X}{dx_X} \approx
   & - \frac{s}{x_X H (1 - \delta_t)} 
 \left(1- \frac{d \log{y}}{d\log{x}} \right)^{-1}
 \langle \sigma v\rangle_{SS\to XX} 
  \frac{(Y_S^{\text{eq}} (T_X))^2 }{ (Y_X^{\text{eq}}(T_X))^2}
  \left(   Y_X^2 - \frac{(Y_X^{\text{eq}} (T_X))^2 }{ (Y_S^{\text{eq}}(T_X))^2} Y_S^2 \right)\,,
       \label{eq:boltz-YX-approx-2}
 \end{align}
 where before freeze-out, due to the sizable effect of  $\langle \sigma v\rangle_{SS\to XX}$,  the following chemical equilibrium is still maintained,
 \begin{align}
\frac{n_X^2}{n_S^2}=  \frac{(n_X^{\text{eq}} (T_X))^2 }{ (n_S^{\text{eq}}(T_X))^2} \,,
 \end{align} 
 with the hidden species companying with the same but nonzero chemical potential. Here, we will use the notations $T_X^f$ and $T_X^c$ for the freeze-out and cannibal decoupling temperatures of the hidden sector particles at their temperatures, respectively, and the corresponding temperatures of the bath are $T^f$ and $T^c$. In Fig.~\ref{fig:relrate}, the green dots approximately denote $m_X/T^c$.
 For $T<T^c$, assuming that $T_X^c \ll m_S -m_X$, the total comoving number density of the hidden sector is approximately conserved, implying,
  \begin{align}
  (n_S +n_X ) a^3 \approx n_X a^3 \approx n_X^{\rm eq} a^3 \Big\vert_{T=T^c} \,,
  \label{eq:comoving-no}
  \end{align}
 due to $n_X \gg n_S$; the result given in Fig.~\ref{fig:relic-boltz1-1} is a suitable example.
Using this above result, at freeze-out temperature, we have
 \begin{align}
 n_X(T_X^f) \approx n_X^{\rm eq} (T_X^c) \frac{a^3(T^c)}{ a^3(T^f)}
 \approx g_X \Big( \frac{m_X T_X^f}{2 \pi} \Big)^{3/2} e^{-m_X/T_X^c} \,,
 \end{align}
 where we have used that after $T<T_c$, $T_X a^2=$ constant, so that $a(T^c)/a(T^f) = (T_X^f / T_X^c)^{1/2}$.
The corresponding yield\footnote{After the cannibal decoupling, the DM evolves with a nonzero chemical potential,
$$\mu = m_X\Big(1 - \frac{T_X}{T_X^c}\Big)\,,$$ which can be obtained using Eq.~(\ref{eq:comoving-no}) and the approximate comoving entropy conservation,
$$ \frac{m_X-\mu}{T_X} n_X a^3 \approx \frac{m_X}{T_X^c}  n_X^{\rm eq} a^3 \Big\vert_{T=T^c} \,.$$
Therefore, $\mu_f =m_X (1 -T_X^f / T_X^c)$  at freeze-out; before that, $X$ and $S$ evolve with the same chemical potential. Actually, the relic abundance is related to $m_X Y_X^\infty \simeq 4.36\times 10^{-10}$\,GeV, corresponding to $\mu |_{T_X \to 0} =m_X$.} of the DM is given by
 \begin{align}
 Y_X(T_X^f) =\frac{n_X(T_X^f)}{s(T^f)} \approx g_X \frac{45}{2^{5/2} \pi^{7/2} h_{\rm eff}} 
 y_f^{3/2} x_f ^{3/2} e^{-x_X^c} 
 & \,,
\left\{
   \begin{array}{lrr}
        \propto e^{-x_X^c}
         \\
         \propto y_f^{3/2}
        \\
        \propto x_f ^{3/2}
         \end{array}
\label{app:gtau}
\right. \;,
 \end{align}
 where $y_f^{3/2} = (T_X^f /T^f)^{3/2}$, $x_f = m_X/T^f$, and $x_X^f = m_X/T_X^f$. The values of $x_X^c$ and $y_f$ (and also for $x_X^f$), relevant to the strength of $3\to2$ interactions and the number densities of $X$ and $S$, can be enlarged by a larger $g_{\rm dm}$ and diminished by the out-of-equilibrium decay of $S$.
 As a result, the much larger the coupling strength $g_{\rm dm}$, the much longer it takes for DM to co-decay with the out-of-equilibrium decay of the mediator, and the lower the DM density to account for the observed relic abundance. 
 
 Thus, we could expect a sizable DM self-interaction cross-section $\sigma_{\rm SI}$ due to having a much larger $g_{\rm dm}$.  
Although the N-body simulation using the collisionless cold DM scenario can successfully describe the Universe's large-scale structure, its predictions are inconsistent with small-scale observations of galaxy formations. Sizable DM self-interactions, which the forbidden dark matter can account for under a cannibally co-decaying phase, alleviate such tensions at a small scale, for instance, core-cusp and too-big-to-fail problems.

\section{Experimental constraints}\label{sec:phenomenology}

The small-scale structure issues that arise from inconsistencies of cold DM $N$-body simulations and observations can be alleviated if sizable self-interactions exist among DM particles \cite{Tulin:2017ara, Adhikari:2022sbh}.  The structure formation related to the too-big-to-fail problem may be resolved by self-interacting DM, resulting in cored halos, with cross sections $\sigma_{\rm SI}/m_X \sim 1$~cm$^2/g$ \cite{Peter:2012jh, Zavala:2012us, Elbert:2014bma}. The halo shapes constrain the self-interacting cross section with a strength $\sigma_{\rm SI}/m_X \lesssim 1 $~cm$^2/g$ \cite{Vogelsberger:2012ku, Peter:2012jh, Zavala:2012us, Elbert:2014bma}.
The halo mergers favor $\sigma_{\rm SI}/m_X \sim$ ($0.1-$few)~cm$^2/g$ \cite{Randall:2008ppe, Harvey:2015hha, Tulin:2017ara, Bradac:2008eu}. Self-interacting DM accommodates the diversity of DM halos with core sizes at the scales of dwarfs and low-surface-brightness galaxies, preferring 
$\sigma_{\rm SI}/m_X \sim$ ($1-10$)~cm$^2/g$ \cite{Dave:2000ar, Elbert:2014bma, Ren:2018jpt, Kaplinghat:2015aga}, and at galaxy-cluster scales, favoring $\sigma_{\rm SI}/m_X \sim$ ($0.1-1$)~cm$^2/g$ \cite{Randall:2008ppe, Kaplinghat:2015aga, Sagunski:2020spe}.
For the small-scale problems, a constant self-interacting cross section with $\sigma_{\rm SI}/m_X \sim 1$~cm$^2/g$ may resolve the core-cusp and too-big-to-fail issues \cite{Tulin:2017ara}. However, to consistently account for different halo masses from dwarf to cluster scales, the self-interacting cross section with a velocity dependence may be preferred \cite{Tulin:2017ara, Kaplinghat:2015aga} for at least one of the dark matter components.

\begin{figure}[t!]
\begin{center}
\includegraphics[width=0.58\textwidth]{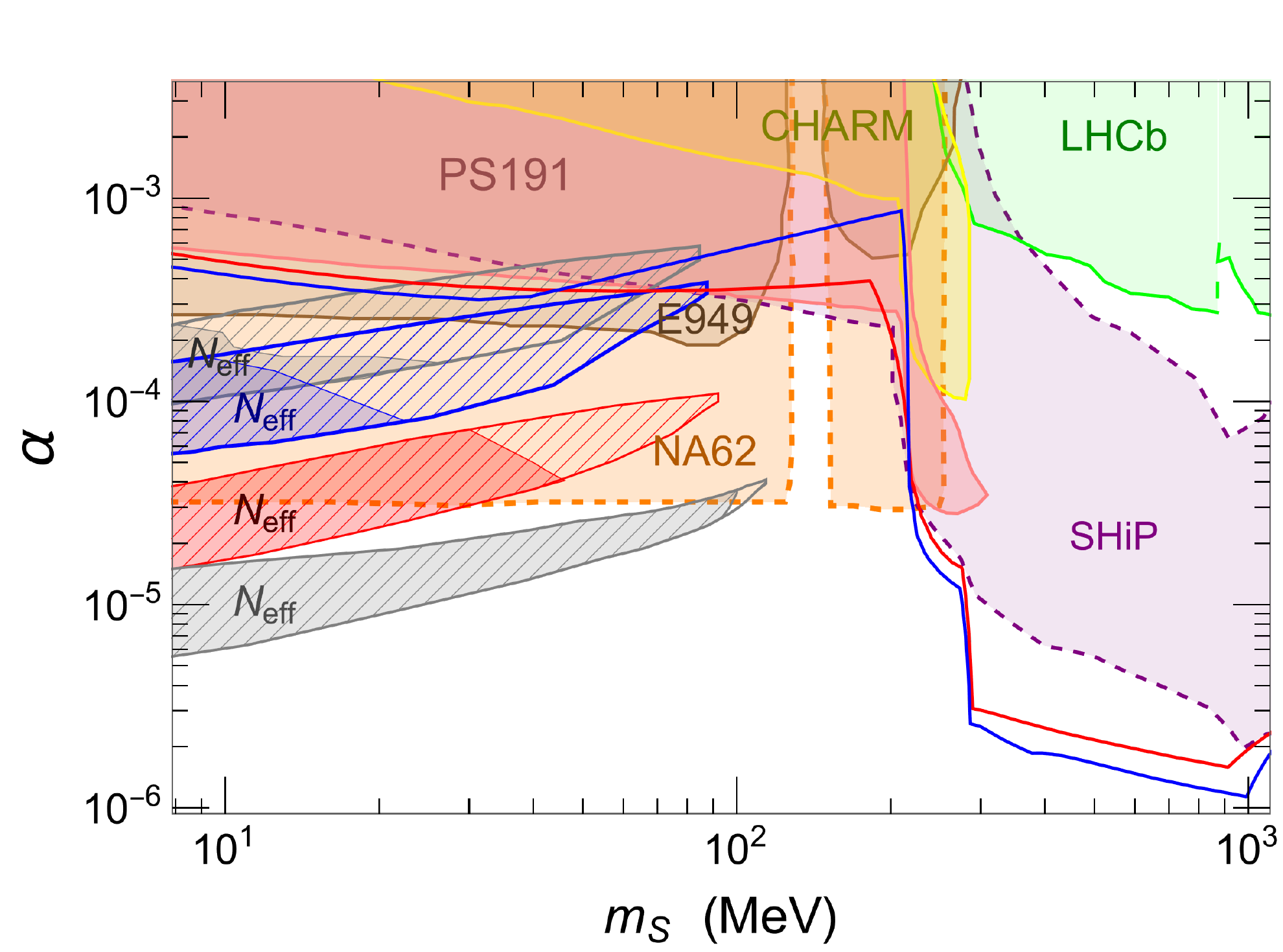}
\caption{Constraints on the $(m_S, \alpha)$ parameter space in the sub-GeV region.
For freeze-out forbidden DM with $r=1.25\, (1.15)$ under a cannibally co-decaying phase, using a value of $g_{\rm dm}$ that leads to observations bounded by $\sigma_{\rm SI}/m_{X} \in (0.1,10)~{\rm cm}^2/g$, the blue (red) hatched area produces the correct relic density, where the shaded region in the same color is excluded by the Planck constraint on $N_{\rm eff}$ at 2$\sigma$. Moreover, the upper (lower) gray region is for $r=1.3\, (1.1)$ for comparison. Colored regions with solid boundaries are experimentally excluded \cite{BNL-E949:2009dza,Bernardi:1985ny, Bernardi:1986hs, Bernardi:1987ek, Gorbunov:2021ccu, CHARM:1985anb, L3:1996ome,Winkler:2018qyg}, while that with dashed boundaries indicate sensitivity projections \cite{Alekhin:2015byh, SHiP:2018yqc, NA62:2017rwk,Bondarenko:2019vrb}. For reference, the solid line (blue for $r=1.25$ or red for $r=1.15$) is for the minimum $\alpha$ of maintaining the hidden sector and bath in thermal equilibrium before freeze-out.
}
\label{fig:constraints-2}
\end{center}
\end{figure}

The present framework gives velocity-independent self-interaction cross-sections in the zero temperature limit.
Using the value of $g_{\rm dm}$ constrained by observations with the strength $\sigma_{\rm SI}/m_{X} \in (0.1,10)~{\rm cm}^2/g$  \cite{Dave:2000ar, Vogelsberger:2012ku, Rocha:2012jg,Peter:2012jh, Zavala:2012us}, we show the regions (hatched) in Fig.~\ref{fig:constraints-2}  for the cases of $r=1.1, 1.15, 1.25$, and $1.3$,  that can account for the correct relic density.  The relevant masses for $X$ and $S$ are $\sim 10-100$~MeV. In the following, we further consider the constraints from particle physics experiments and astrophysical and cosmological observations.

\subsection{The flavor and fixed target experiments}

\begin{figure*}[t!]
\begin{center}
\vskip-0.35cm
\includegraphics[width=0.45\textwidth]{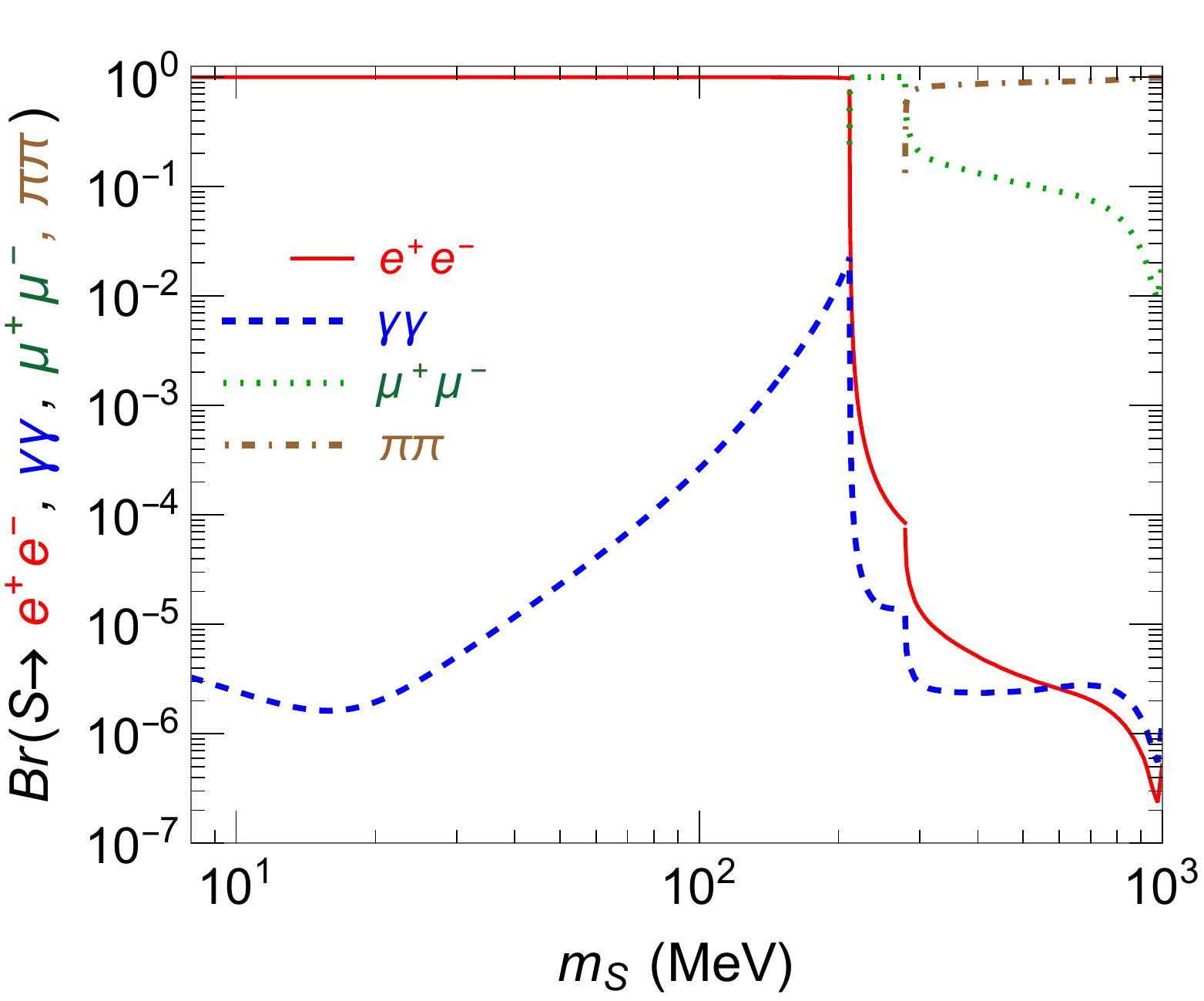}\hskip0.5cm
\includegraphics[width=0.46\textwidth]{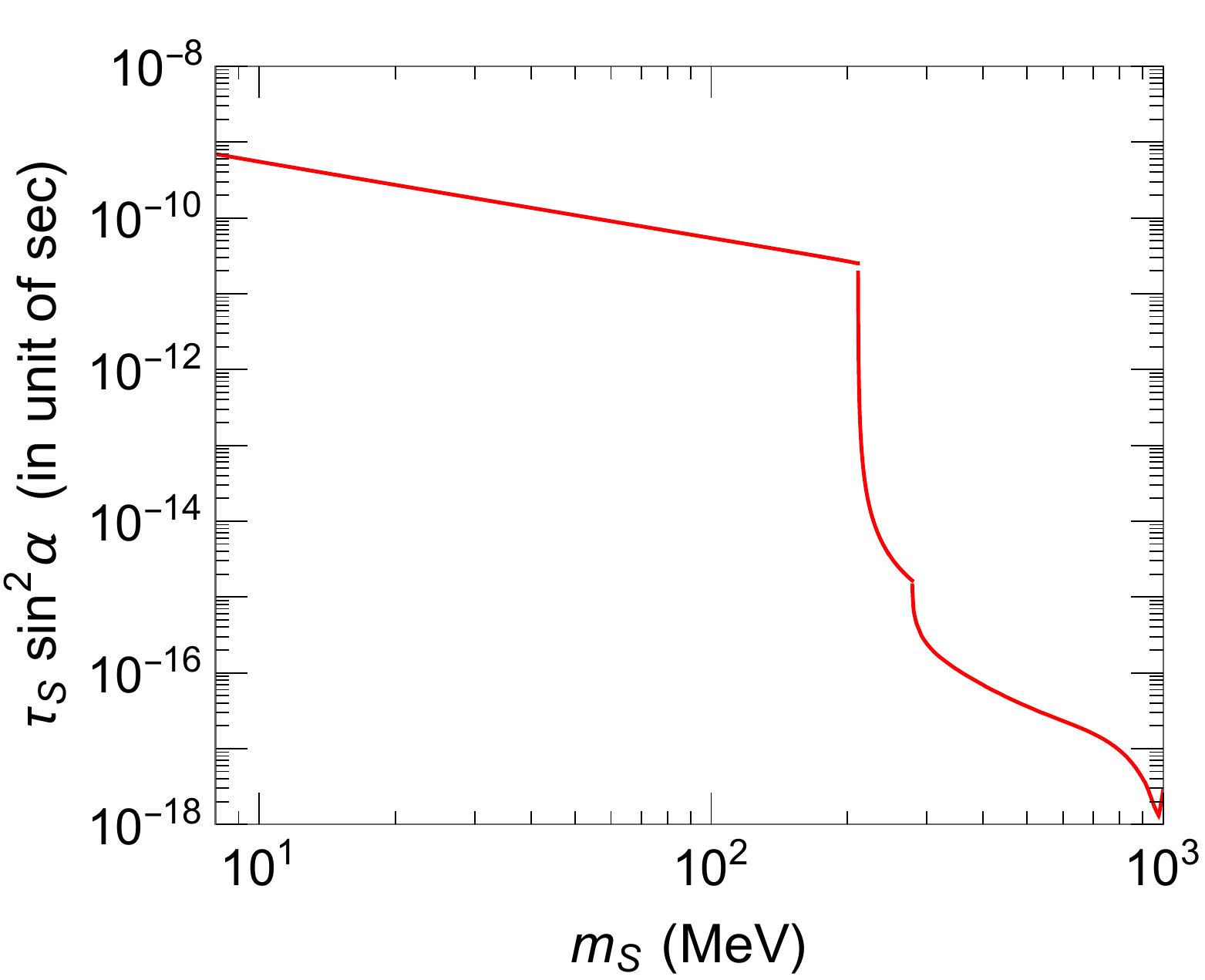}\\
\vskip-0.35cm
\caption{Left panel: Branching ratios of the hidden scalar decays as functions of $m_S$ in the range of 7~MeV$-$1000~MeV.  Right panel: Hidden scalar lifetime ($\tau_S$) vs. $m_S$, where $\tau_S$ is shown by multiplying a factor of $\sin^2\alpha$.}
\label{fig:br-life}
\end{center}
\end{figure*}

By mixing with the SM Higgs, the unstable light scalar, which resides within a hidden sector and is the partner of the forbidden DM, can couple with the SM particles. Therefore,  this sub-GeV hidden scalar could be produced in rare decays performed at colliders and fixed target experiments. The branching ratios ($Brs$) and lifetime ($\tau_S$) of the hidden scalar $S$ as functions of $m_S$ are shown in Fig.~\ref{fig:br-life}, where $\tau_S\propto 1/\sin^2\alpha$.
As for the monojet + MET (missing transverse energy) search at the LHC colliders, the direct production rate $p p (\to j + S^*) \to j + X X $, involving the DM particles in the final state, is highly suppressed owing to the slight mixing angle $\alpha$ and $m_S>m_X$ in this case.
Instead, the events containing the on-shell $S$ in the final states are conceivably measurable, for which searches for missing energy signals, resulting from the long-lived scalar $S$ decays outside the detectors, or displaced decays from beam dump experiments become relevant \cite{BNL-E949:2009dza, Bernardi:1985ny, Bernardi:1986hs, Bernardi:1987ek, Gorbunov:2021ccu, CHARM:1985anb, L3:1996ome, Winkler:2018qyg, Alekhin:2015byh, SHiP:2018yqc, NA62:2017rwk, Bondarenko:2019vrb}.

Because its couplings to matter are relatively suppressed by the smallness of $\sin\alpha$, searches for missing energy signals, for which the long-lived scalar $S$ decays outside the detectors, or displaced decays from beam dump experiments become relevant \cite{BNL-E949:2009dza, Bernardi:1985ny, Bernardi:1986hs, Bernardi:1987ek, Gorbunov:2021ccu, CHARM:1985anb, L3:1996ome, Winkler:2018qyg, Alekhin:2015byh, SHiP:2018yqc, NA62:2017rwk, Bondarenko:2019vrb}.

Fig.~\ref{fig:constraints-2} shows the experimental constraints on the parameter space of the $(m_S,\alpha)$ plane, where the relevant $m_S$ is in the sub-GeV region.  
As shown in Fig.~\ref{fig:constraints-2}, finding the long-lived scalar particle at beam dump experiments with detectors far from the event generation is promising.
The PS191 measurement \cite{Bernardi:1985ny, Bernardi:1986hs, Bernardi:1987ek}, operated in the '80s, was recently re-analyzed to search for the light scalar decaying into charged particles by taking into account the process $K\to \pi S$ \cite{Gorbunov:2021ccu}.
The data of searching for $K^+ \to \pi^+ \bar{\nu} \nu$ by E949 \cite{BNL-E949:2009dza} can be used to set bounds on the branching ratio of $K^+ \to \pi^+$ together with a long-lived scalar $S$, which escapes detection. Currently, the PS191 and E949 experiments put the upper limit on $\alpha$ in between $2\times 10^{-4} - 6\times 10^{-4}$ in the relevant $m_S$ region. 

The upcoming run of NA62 \cite{NA62:2017rwk} is at the CERN SPS. One of the measurements in NA62 is the rare decay $K^+ \to \pi^+ \nu \bar\nu$, which displays as $\pi^+ +$ missing energy in the final state. If the scalar $S$ is long-lived enough to decay outside the detector, searching for $K^+ \to \pi^+ +$ missing energy can set constraints on the $(m_S, \alpha)$ parameter plane.
 In Fig.~\ref{fig:constraints-2}, we depict the region where the future sensitivity of NA62 can reach after LHC Run 3 \cite{Bondarenko:2019vrb}.
 NA62 is expected to improve the constraint on $\alpha$ by a factor of $7-10$ compared with E949.
 It is likely to discover the signal hint for the self-interacting forbidden DM through the future NA62 beam dump experiment.

\subsection{Number of relativistic degrees of freedom, $N_{\rm eff}$}

In the mass range of interest ({\it e.g.,} the hatched regions in Fig.~\ref{fig:constraints-2}), the DM freeze-out occurs after the neutrinos have decoupled with temperature $T_{\rm dec}\sim 1$~MeV (see also Fig.~\ref{fig:relic-boltz1-1}).
Meanwhile, after neutrino decoupling, the produced particles, which are majorly electron and positron for 8~MeV$< m_S< 2m_\mu$, in the out-of-equilibrium decay of $S$ can be rapidly thermalized with the background photons. See Fig.~\ref{fig:br-life} for reference. As such, considering the mass range  8~MeV$< m_S< 2m_\mu$, that we are interested in, the produced electrons and positrons are finally Boltzmann suppressed, and the most energy injected from the hidden sector via $S$ out-of-equilibrium decay contributes to the background photons.
 
 Thus, after neutrino decoupling, the background photon energy density is approximately changed as $\rho_\gamma \to \rho_\gamma + \delta\rho_\gamma$ with $\delta\rho_\gamma \approx  (\rho_S+ \rho_X)_{T=T_{\rm dec}}$ and $\rho_S+ \rho_X \simeq m_S n_S + n_X n_X$. Here, $\rho_\gamma = (\pi^2/15) \tilde{T}^4$ with $ \tilde{T}=T_\nu (11/4)^{1/3}$ and $T_\nu$ being the neutrino temperature.
In other words, the photon bath is further heated relative to neutrinos, thereby reducing the effective extra number of relativistic species,
\begin{align}
N_{\rm  eff} =  \frac{8}{7} \left(\frac{11}{4} \right)^{4/3}
 \left. \frac{3\rho_\nu}{\rho_\gamma + \delta\rho_\gamma} \right\vert_{T=T_{\rm dec}} \,,
\end{align}
where $\rho_\nu = (7/8) (\pi^2/15) T_\nu^4 $ is the energy density of a species of neutrino.
Combining with baryon acoustic oscillation (BAO), Planck 2018 constrains $N_{\rm eff} = 2.99 \pm 0.17$ \cite{Planck:2018vyg}. Using the SM value $N_{\rm eff}^{\rm SM} =3.046$ \cite{Mangano:2005cc}, we limit $\Delta N_{\rm eff} = N_{\rm eff} - N_{\rm eff}^{\rm SM} \gtrsim -0.396$ at 2$\sigma$, resulting in the requirement of $m_S Y_S +m_XY_X \lesssim 1.9\times 10^{-5}$~GeV at $T_{\rm dec}=1~{\rm MeV}$. This constraint, shown in Fig.~\ref{fig:constraints-2}, can rule out some parameter regions with $m_S\lesssim 9$~MeV. 

$\Delta N_{\rm eff}$ and E949 can further constrain the mass ratio to lie in the range $1.1\lesssim r (=m_S/m_X) \lesssim 1.33$ and the scalar mass to be $ 9~{\rm MeV} \lesssim m_S \lesssim 114$~MeV.
 Moreover, the nearly favored parameter space is projected to be testable with the NA62 experiment.

\subsection{Big Bang Nucleosynthesis (BBN) }

Ref.~\cite{Hufnagel:2018bjp} has recently updated the BBN constraints on a MeV Higgs-like scalar\footnote{Both $\phi$ and the hidden $S$ are Higgs-like scalars. In Ref.~\cite{Hufnagel:2018bjp}, the authors discussed $T_\phi^{\rm cd}/T^{\rm cd} >1$ in the relativistic region. Nevertheless, for our case, although  $T_\phi^{\rm cd}/T^{\rm cd} >1$, the relevant processes occur in the nonrelativistic region.} $\phi$, which is chemically decoupled from its hidden sector partners at its temperature $T_\phi^{\rm cd}$, corresponding to photon temperature $T^{\rm cd}$. Taking the condition $T_\phi^{\rm cd} =T^{\rm cd} \gg m_\phi$ as a benchmark choice, they obtained the upper bound on the lifetime $\tau_\phi\approx 0.3\, {\rm s}$ for $m_\phi=100$~MeV, but becomes 10\,s for a smaller $m_\phi=10$~MeV for $Br(S\to e^+ e^-)=1$ (see the left panel of Fig.~5 in Ref.~\cite{Hufnagel:2018bjp}). For our case, the scalar $S$ is chemically decoupled from the DM until freeze-out, which may occur at temperatures less than 1~MeV (the neutrino decoupling temperature). 

Therefore we estimate the BBN constraints by comparing the residue percentage of the comoving energy density, that may then inject into the bath for $T<1$~MeV,
$a= [(m_X Y_X + m_S Y_S) h_{\rm eff}]|_{T=1\,{\rm MeV}} / (m_S Y_S  h_{\rm eff}) |_{T=10\,{\rm GeV}}$\footnote{We have used $n_\gamma \propto s / h_{\rm eff} $. For $S$ being relativistic, its moving number density keeps~ constant. For nonrelativistic $S$ that could decay during temperatures $T_{\rm dec}  >T > T_f$, unlike Ref.~\cite{Hufnagel:2018bjp}, we need to take into account the extra comoving energy density generated from $XX\to SS$, which is proportional to $m_X Y_X  h_{\rm eff}|_{T=1\,{\rm MeV}}$.} in our case to $b= (n_\phi /n_\gamma )|_{ T=(1\,{\rm MeV})}/ (n_\phi /n_\gamma)|_{T=10\,{\rm GeV}}$ 
with $(n_\phi /n_\gamma)|_{T=10\,{\rm GeV}}=1/2$ and $n_\gamma$ being the photon number density, where the upper bound of $b$ can be obtained since $\tau_\phi$ is constrained by BBN given in Ref.~\cite{Hufnagel:2018bjp}. If $a>b$, the result is excluded by BBN; otherwise, it is allowed. For our case, the exclusion also depends on $r$ and $g_{\rm dm}$.  Below are some examples that provide the correct relic density and account for small-scale problems.
For instance, for $m_S=25$~MeV, $b\lesssim 0.65$ is constrained by the BBN bound, compared with $a\simeq 4\times 10^{-4}$ (allowed) corresponding to the case with $r=1.25, g_{\rm dm}=1.083$, and $\alpha=2.45\times 10^{-4}$ (the case shown in Fig.~\ref{fig:relic-boltz1-1}), or $a \approx 1.4$ (excluded) corresponding to the case with $r=1.1, g_{\rm dm}=1.09$, and $\alpha=1.95\times 10^{-5}$; for $m_S=70$~MeV, the BBN constraint gives $b\lesssim 0.21$, compared with $a \approx 10^{-6}$ (allowed) corresponding to the case with $r=1.25, g_{\rm dm}=2.77$, and $\alpha=3.84\times 10^{-4}$, or $a \approx 0.18$ (allowed) corresponding to the case with $r=1.1, g_{\rm dm}=2.36$, and $\alpha=2.84\times 10^{-5}$. 

We do not further show the BBN constraint in the plot because we find that it is generally comparable but weaker than that from $N_{\rm eff}$; this is consistent with the result for $m_\phi \gtrsim 2$~MeV shown in the left panel of Fig.~5 in Ref.~\cite{Hufnagel:2018bjp}.

\section{Conclusions}\label{sec:conclusions}

 In the present scenario, we have considered that a Higgs-like scalar $S$ is contained in the hidden sector. Thus, the hidden sector particles interact with the SM through the mixture of the SM-like Higgs and hidden Higgs. In the study, we adopt the simplest vector DM model, where the hidden sector's DM ($X$) is an abelian gauge vector boson but with $m_X < m_S$.
 The thermally averaged forbidden rate of the two-body DM annihilation, $XX\to SS$, is exponentially suppressed compared with its inverse process.
 
If the hidden sector is in chemical and kinetic equilibrium with the bath before freeze-out, a much larger mass ratio $m_S/m_X \in (1.5,1.65)$ will be required to simultaneously provide the correct relic density and proper size of DM self-interactions for accounting for the small scale problems. However, for this condition,
the mixing angle between the SM-like Higgs and hidden scalar, corresponding to the viable parameter region, is ruled by E949, PS191, and CHARM experiments.

We have found that the region of the parameter space with a smaller $\alpha$,  where the hidden sector is thermally decoupled from the SM at $T\sim m_X$, can result in $g_{\rm dm} \propto {\cal O}(1)$, from which the correct DM relic density and proper size of DM self-interactions can thus be obtained. 
For the underlying mechanism of this scenario, 
 when the hidden sector is decoupled from the SM at $T\sim m_X$, it enters a cannibally co-decaying phase and evolves with an independent temperature.
 As such, a much larger coupling strength ($g_{\rm dm}$) between $X$ and $S$ is needed to have a significantly longer interacting time for them to deplete the DM density with the out-of-equilibrium decay of $S$ so that the value of relic abundance can be reduced to be consistent with observation.

We have shown that a sizable parameter space still survives the most current constraints. The favored region, constrained mainly by Planck $N_{\rm eff}$ and E949, has the mass ratio $1.1\lesssim r (=m_S/m_X) \lesssim 1.33$ and the scalar mass $ 9~{\rm MeV} \lesssim m_S \lesssim 114$~MeV.
The projected sensitivity of the NA62 beam dump experiment can further probe the parameter space of the hidden scalar.


\begin{acknowledgments}
 This work was partly supported by the National Center for Theoretical Sciences and the National Science and Technology Council of Taiwan under grant numbers, MOST 111-2112-M-033-006 and NSTC 112-2112-M-033-007.
\end{acknowledgments}


\appendix

\section{The thermal freeze-out}\label{thermal_freezeout}

Here, we consider the condition that the hidden sector particles are in thermal equilibrium among themselves and with the SM bath during the freeze-out process, which also infers $T_i\equiv T_X =T$. Thus, from Eq.~(\ref{eq:boltz-11}), the evolution of the DM number density can be approximately described by
\begin{align}
\frac{d Y_X}{dx} \approx
   & - \frac{s \langle \sigma v\rangle_{SS\to XX}   }{x H} 
 \frac{(Y_S^{\text{eq}} (T) )^2 }{ (Y_X^{\text{eq}} (T) )^2}
  \left[   Y_X^2 - \big(Y_X^{\rm eq} (T) \big)^2 \right]  \,.
  \label{eq:boltz-YX-approx-1}
 \end{align}
 Analogous to calculating the WIMP (weak interacting massive particle) DM, one can obtain the following results for the s-wave $S S \leftrightarrow XX$ annihilation. 
\begin{align}
 \frac{x_f^{1\over 2} e^{(2r-1)x_f}  } { \langle \sigma v \rangle_{SS\to XX} }
 &\approx
\delta (\delta+2) \sqrt{\frac{45}{4 \pi^5}} \frac{M_{\rm pl} m_X r^3}{ g_{\rm eff}^{1/2}(T_f)  } \frac{g_S^2 }{g_X} \,,
\label{eq:xf} \\
%
\frac{1}{m_X Y_X^{\infty}}
& \approx
\frac{4\pi}{\sqrt{90}}M_{\rm pl} g_*^{1/2}(T_f)
 \langle \sigma v\rangle_{SS\to XX}
 \frac{g_S^2 }{g_X^2}  r^3 \nonumber\\
 &\times \bigg\{  \frac{e^{-\Delta_f}}{x_f}  +   \frac{\Delta_f}{x_f}    {\rm E_i} (-\Delta_f)
  -\frac{15}{8}\Big( 1-\frac{1}{r} \Big)
\bigg[ \frac{1-\Delta_f}{x_f^2} e^{-\Delta_f}  - \frac{\Delta_f^2}{x_f^2}  {\rm E_i} (-\Delta_f)
    \bigg] \bigg\} \,,
 \label{eq:xs}
\end{align}
where $g_{\rm eff}$ is the effective DoFs of the Universe's total energy density,  $g_*^{1/2} \equiv \tilde{h}_{\rm eff}  /g_{\rm eff}^{1/2}$,
$M_{\rm pl} \equiv (8\pi G)^{-1/2}$ is the reduced Planck mass, we use $\delta = 1$ defined by $Y_X - Y_X^{\rm eq} = \delta Y_X^{\rm eq}$ with $d\delta/dx \ll 1$,  $\Delta_f\equiv 2(r-1)x_f$,
${\rm Ei}(z)= - \int_{-z}^{\infty} (e^{-t}/t) dt$, and $ Y_X^{\infty}$ is the  post-freeze-out value of the yield, related to the DM relic abundance \cite{pdg2022}
$\Omega_X =  m_X Y_X^\infty  s_0 /\rho_c  \simeq 0.1198/h^2 $ with the present critical density
$\rho_c = 1.0537\times 10^{-5} h^2 ({\rm GeV}/c^2) {\rm cm}^{-3}$, $h\simeq 0.674$, and the present entropy $s_0=2891~\text{cm}^{-3}$.

\begin{figure*}[t!]
\begin{center}
\includegraphics[width=0.62\textwidth]{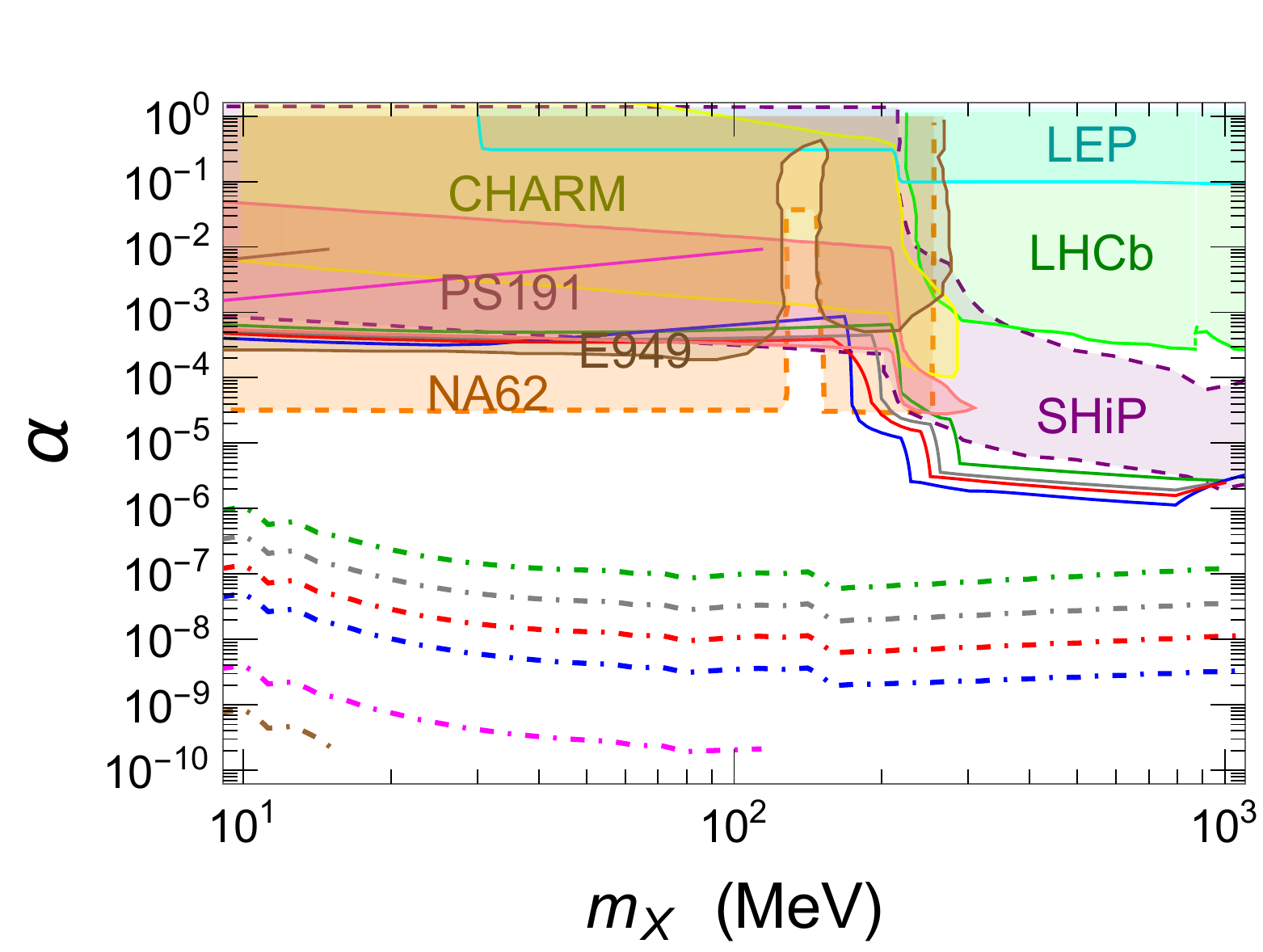}
\caption{ (a) The solid lines show the minimum value of $\alpha$ about keeping the hidden sector in thermal equilibrium with the bath before freeze-out.
(b) Dotdashed lines show the minimum $\alpha$ above that $Y_{X, S}$ grow by up to $Y_{X, S}^{\rm eq}(T)$ at $T \gtrsim m_X$ by a freeze-in process. (c) colored regions with solid boundaries are excluded by experiments denoted  \cite{BNL-E949:2009dza, Gorbunov:2021ccu, CHARM:1985anb, L3:1996ome,Winkler:2018qyg}, while that with dashed boundaries indicate sensitivity projections  \cite{Alekhin:2015byh, SHiP:2018yqc, NA62:2017rwk,Bondarenko:2019vrb}, where for simplicity $r=1.01$ is used. The green, gray, red, blue, magenta, and brown lines correspond to $r (=m_S/m_X)=1.01, 1.1, 1.15, 1.25, 1.5$, and 1.65, respectively.}
\label{fig:model-xs-mx-gdm}
\end{center}
\end{figure*}

Using the model result for $SS\to XX$ annihilation cross section as a function of the coupling strength $g_{\rm dm }$
 and determining the value of $m_X Y_X^{\infty}$ from fitting to the observed relic density, Eqs.~(\ref{eq:xf}) and (\ref{eq:xs}) numerically give $g_{\rm dm }$ and $x_f$ are functions of $m_X$, with respect to a different value of $r$. The result for $g_{\rm dm }$ has been shown in Fig.~\ref{fig:model-mx-gdm}. 
 
  In Fig.~\ref{fig:model-xs-mx-gdm}, we further display the minimum value of $\alpha$ that is required to keep the hidden sector and the bath in the thermal equilibrium before freeze-out\footnote{Compared with that shown in Ref.~\cite{Yang:2022zlh}, in addition to a curve with $r=1.65$ added, numerical errors relevant to the MeV region are corrected in the present work.}, where the reactions are via elastic ($X \,  {\rm SM} \leftrightarrow X \,  {\rm SM}$ and $S \,  {\rm SM} \leftrightarrow S \,  {\rm SM}$) and number-changing ($S \leftrightarrow {\rm SM}~{\rm SM}$) interactions. Numerically, we have checked that if $\alpha$ is more prominent than its minimum, then the correct relic abundance is highly insensitive to it, as expected. 
 As shown in Fig.~\ref{fig:model-mx-gdm}, only a much larger mass ratio $r\in (1.5,1.65)$ can simultaneously account for the correct relic density and the requirement of the small-scale problems.
  Nevertheless, the corresponding parameter space\footnote{For $r=1.5\, (1.65)$, the parameter space is restricted in the region $m_X \lesssim 114\, (15)$~MeV, owing to the unitarity bound is imposed to $g_{\rm dm}$ (see Fig.~\ref{fig:model-mx-gdm}).}, required by $\alpha>10^{-3}$ for $m_X>10$~MeV, is excluded by E949 \cite{BNL-E949:2009dza}, PS191 \cite{Bernardi:1985ny, Bernardi:1986hs, Bernardi:1987ek, Gorbunov:2021ccu}, and CHARM \cite{CHARM:1985anb} measurements.

For references, the experimental bounds are summarized in Fig.~\ref{fig:model-xs-mx-gdm}. 
We also show the minimum $\alpha$ (dot-dashed lines)\footnote{Compared with Ref.~\cite{Yang:2022zlh}, numerical errors especially relevant to the MeV region are corrected, too.}, assuming $S$ is generated through a freeze-in process so that $Y_S$ can reach the value of $Y_S^{\rm eq}$ at $T=m_X$, and, meanwhile, $X$ is quickly thermalized via $XX\leftrightarrow SS, XS \leftrightarrow XS$. This is estimated by
\begin{align}
\int_0^{Y_S^{\rm eq}(m_X) }dY_S \simeq
   \int_{0}^{1} \frac{\tilde{h}_{\rm eff}}{h_{\rm eff}}  
   \frac{ s \langle \sigma v \rangle_{SS\to {\rm SM}\, {\rm SM}} Y_S^{\text{eq}}{\big.}^2
    +
  \langle \Gamma_{S} \rangle  Y_S^{\text{eq}}
   } {xH}
    dx \,,
   \label{eq:freezein}
\end{align}
where the value shown in Fig.~\ref{fig:model-mx-gdm} for $g_{\rm dm}$, relevant to the coupling $g_{SSS}\sim -3 \cos^3\alpha\, m_S^2 g_{\rm dm}/m_X$ \cite{Yang:2022zlh} for calculating the $SS\to {\rm SM}\, {\rm SM}$ cross section, has been used.
As shown in Fig.~\ref{fig:relrate} that $\langle \sigma v \rangle_{XX\to {\rm SM}\, {\rm SM}} Y_X^{\text{eq}}{\big.}^2 \sim \langle \sigma v \rangle_{SS\to {\rm SM}\, {\rm SM}} Y_S^{\text{eq}}{\big.}^2 $, the estimate result for the freeze-in of DM will be similar to $S$. Because the allowed $\alpha$ relevant to the present study (see Fig.~\ref{fig:constraints-2}) is above the dot-dashed line of Fig.~\ref{fig:model-xs-mx-gdm}, here we thus do not further include contribution arising from DM freeze-in.

\section{The case of Majorana dark matter}\label{app:majorana}

\begin{figure*}[t!]
\begin{center}
\includegraphics[width=0.68\textwidth]{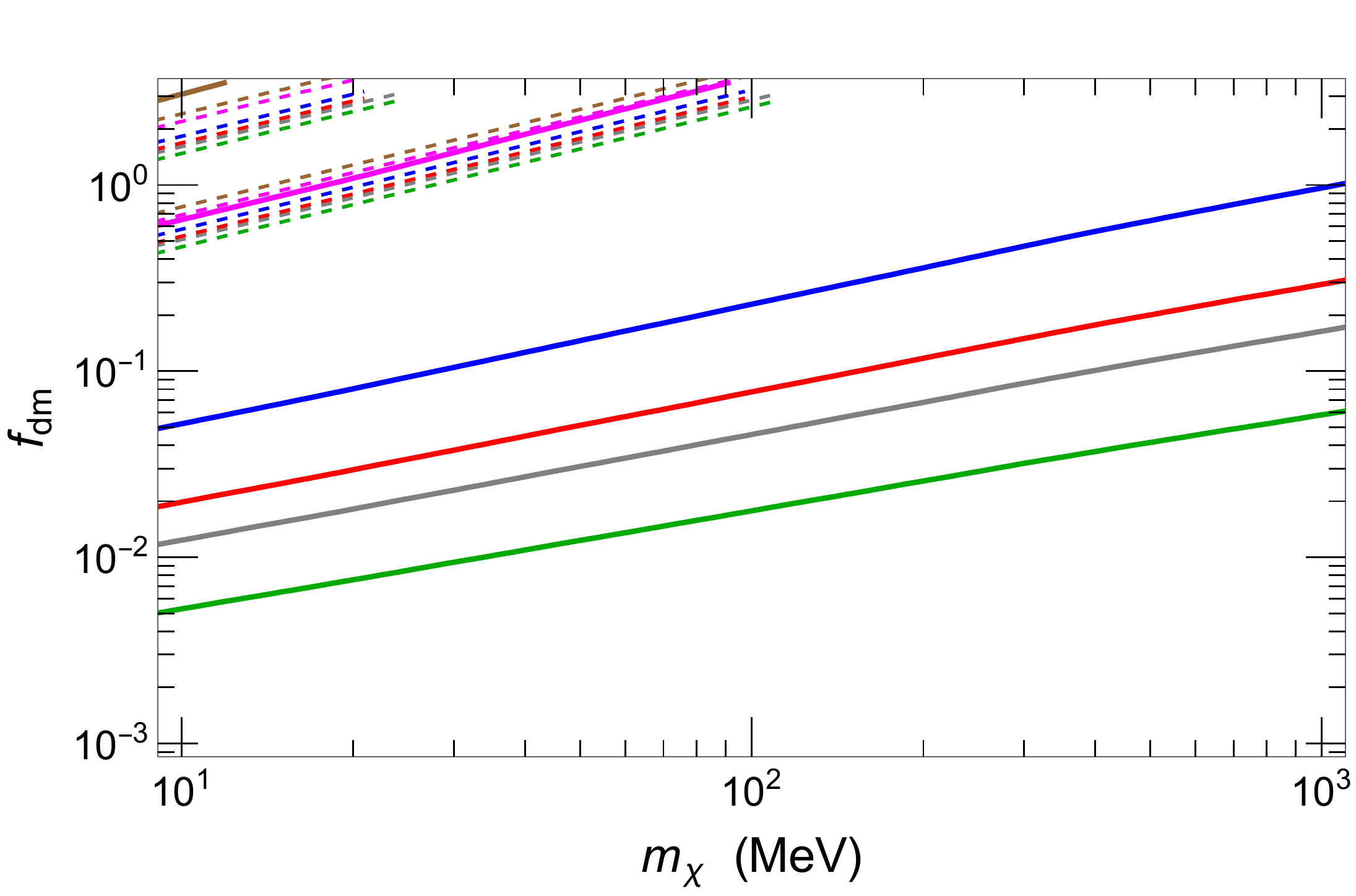}
\caption{$f_{\rm dm}$ as a function of $m_\chi$, where $\lambda_3$ is taken to be zero. The rest are the same as Fig.~\ref{fig:model-mx-gdm}.}
\label{fig:model-mx-fdm}
\end{center}
\end{figure*}

In the main body of this paper, we study that the hidden sector is composed of a vector DM boson and a scalar. Similarly, our present scenario also works if a DM candidate is a Majorana fermion $\chi$. As an example, we consider that the relevant Lagrangian, where the hidden sector involves $\chi$ and a scalar $S$, can be described by \cite{Farina:2016llk}
\begin{align}
{\cal L} \supset  &\frac{i}{2} \bar\chi \gamma_\mu \partial^\mu \chi   
                        -\frac{1}{2}  m_\chi \bar\chi \chi 
                         -\frac{f_{\rm dm}}{2} S \bar\chi \chi  +\frac{1}{2} \partial^\mu S \partial_\mu S - V(S,h) \,.
 \label{eq:L-DS-DMr}
\end{align}
We assume that the scalar $S$ results from another scalar that weakly mixes with the SM-like Higgs $h$ due to spontaneous symmetry breaking so that
\begin{align}
V(S, h) = \frac{m_S^2}{2} S^2 + \frac{\lambda_3}{3!} S^3 + \frac{\lambda_4}{4!} S^4 + \cdots\,.
\end{align}
This simple model characterizes the main feature of Weinberg’s Higgs portal model with a slight mixing angle for the Higgs sectors \cite{Weinberg:2013kea}. The Weinberg model further contains a Goldstone boson which can be constrained by $N_{\rm eff}$.

For the forbidden Majorana model, the process is mainly related to the $s$-wave $ S S \to \chi \chi$ annihilation with the cross-section (with $c _\alpha f_{\rm dm}$ replacing $f_{\rm dm}$ in Eq.~(\ref{eq:L-DS-DMr})),
\begin{align}
& (\sigma v_\text{lab})_{SS\to \chi \chi} =
\nonumber\\
& c_\alpha^2 f_{\rm dm}^2 
\Bigg\{ 
2 \left( m_S^4 + m_f^2 (s - 4 m_S^2 )\right) 
\Big[
c_\alpha^2 f_{\rm dm}^2  \left(\Gamma_S^2 m_S^2+ (s- m_S^2)^2\right) \left( 2 m_S^4 - 4 s m_S^2 + s^2+m_f^2 (4s -8 m_S^2 )\right)
\nonumber\\
& - 2 c_\alpha f_{\rm dm} \lambda_3 m_f  (s- 4 m_f^2 ) \left(2 m_S^4-3 s m_S^2+s^2\right)
\Big] 
\log \Bigg( \frac{ s-2 m_S^2 + \sqrt{s-4 m_f^2} \sqrt{s-4 m_S^2} } {  s- 2 m_S^2 - \sqrt{s-4 m_f^2} \sqrt{s-4 m_S^2} } \Bigg)
\nonumber\\
&  -\sqrt{s-4 m_f^2} \sqrt{s-4 m_S^2} (s-2 m_S^2) \nonumber\\
&\ \ \ \times \Big[2 c_\alpha^2 f_{\rm dm}^2 (8 m_f^4-4 m_S^2 m_f^2+m_S^4 ) \left(\Gamma_S^2 m_S^2+ (s- m_S^2)^2\right) -  \lambda_3^2 (s- 4 m_f^2) \left(m_S^4+m_f^2  (s-4 m_S^2 )\right) \Big]
\Bigg\} 
\nonumber\\
&\times \frac{1}{16 \pi  \left(\Gamma_S^2 m_S^2+\left(m_S^2-s\right)^2\right) \sqrt{s} \sqrt{s-4 m_S^2} \left(s-2 m_S^2\right)^2 \left(m_S^4+m_f^2 (s-4 m_S^2)\right)} \\
& \to \frac{ c_\alpha^2 \,  f_{\rm dm}^2 (m_S^2 - m_\chi^2)^{3/2} 
\Big( 4 c_\alpha^2 \,  f_{\rm dm}^2  \big( 9 m_S^2 + \Gamma_S^2 \big)   m_\chi^2 -12 c_\alpha\,  f_{\rm dm} \lambda_3 m_\chi m_S^2 + \lambda_3^2 m_S^2 \Big)}{8 \pi m_S^7 \big( 9 m_S^2 + \Gamma_S^2 \big)} + {\cal O}(v_{\rm lab}^2) \,,
\end{align}
where $c_\alpha\equiv \cos\alpha$, $m_S > m_\chi$, and $v_{\rm lab}$ is the relative velocity in the rest frame of one of the collision particles.
In the Higgs portal model,  the coupling $\lambda_3 \sim -3 \cos^3\alpha\, m_S^2/v_S$.
Under the condition that the hidden sector particles are not only in thermal equilibrium among themselves but also with the SM bath during freeze-out, and the correct relic density can be well produced, we can evaluate, using the formulas given in Appendix~\ref{thermal_freezeout}, the value of $f$ as a function of $m_\chi$. The result is shown in Fig.~\ref{fig:model-mx-fdm}, where we have taken $\lambda_3=0$  and $\cos\alpha=1$.

For the self-interacting DM scattering $\chi \chi \to \chi \chi$ through the scalar mediator exchange via $u$- and $t$-channels can be obtained in the zero-velocity limit, given by (again with $c _\alpha f_{\rm dm}$ replacing $f_{\rm dm}$ in Eq.~(\ref{eq:L-DS-DMr}))
\begin{align}
\sigma_{\rm SI} = 
\frac{ c_\alpha^4 \,  f_{\rm dm}^4 m_\chi^2 }{32 \pi  m_S^4}\,.
\end{align}
Constraining $g_{\rm dm}$ to account for the strength $\sigma_{\rm SI}/m_{X} \in (0.1,10)~{\rm cm}^2/g$, we also show the allowed value in Fig.~\ref{fig:model-mx-fdm}. The resultant $f_{\rm dm}$ is order of one but becomes slightly $r$-dependent. The Majorana case's corresponding properties are consistent with the vector DM one (see Fig.~\ref{fig:model-mx-gdm}).

Moreover, if the forbidden Majorana DM (together with the hidden scalar) is decoupled from the SM bath at $T\sim T_X$ and then undergoes a cannibally co-decaying phase, the $3\to 2$ rates should be larger than the cosmic expansion rate almost until when the temperature drops close to $T_f$. The relevant $3 \to 2$ cannibalizations include $SSS \to \chi \chi, \chi\chi\chi \to \chi S, \chi \chi S \to SS, \chi SS \to \chi S$ and $SSS\to SS$. Except for the last one, the others involve the term proportional to $f_{\rm dm}^6$, while all of the corresponding ones in the vector DM case are in the same order of magnitudes. Here, we thus give $SSS \to \chi \chi$,
\begin{align}
\langle \sigma v^2 \rangle_{SSS\to \chi\chi} & \simeq \frac{(9 m_S^2 - 4m_\chi^2)^{3/2}}{12288 \pi m_S^{12} }  f_{\rm dm}^2 
 \big[ m_S^2 (\lambda_4 -12 f_{\rm dm}^2) + (\lambda_3  -12 f_{\rm dm} m_\chi ) (\lambda_3 - 4f_{\rm dm} m_\chi)\big]^2 \,,
  \end{align}
  which appears in the DM number density evolution as
 \begin{align}
\frac{dn_\chi}{dt} + 3 H n_\chi=
& \cdots 
+ \frac{1}{3} \langle \sigma v^2 \rangle_{SSS\to \chi\chi} \bigg( n_S^3 -(n_S^{\text{eq}})^3 \frac{n_\chi^2}{(n_\chi^{\text{eq}})^2} \bigg)  \nonumber\\
& -  \langle \sigma v^2 \rangle_{\chi\chi S\to SS} \bigg( n_\chi^2 n_S -(n_\chi^{\text{eq}})^2 \frac{n_S^2}{(n_S^{\text{eq}})} \bigg)
    -  \frac{1}{3} \langle \sigma v^2 \rangle_{\chi\chi\chi\to \chi S} \bigg( n_\chi^3 -n_\chi n_S \frac{(n_\chi^{\text{eq}})^2}{n_S^{\text{eq}}} \bigg)   
   \,, \label{eq:boltz-1-majorana}
\end{align}
 to compared with  $SSS \to XX$ in the vector DM case,
  \begin{align}
 \langle \sigma v^2 \rangle_{SSS\to XX} & \simeq \frac{\sqrt{9-4m_X^2/m_S^2}}{384 \pi m_S^3 m_X^2}  g_{\rm dm}^6 m_X^2
 \nonumber\\
  & \times \Bigg(  \frac {432  m_X^6} {m_S^8} - \frac {2160 m_X^4} {m_S^6} + \frac {729  m_S^4} {16 m_X^6} 
  + \frac {3672  m_X^2} {m_S^4} + \frac {891   m_S^2} {4 m_X^4} - \frac {1944 } {m_S^2} - \frac {729  } {4 m_X^2}
  \Bigg) \,.
  \end{align}
By simply taking $m_\chi =m_S=m_X$ and using $\lambda_3=\lambda_4=0$, we find that the ratio is about 
\begin{align}
\frac{\langle \sigma v^2 \rangle_{SSS\to \chi\chi} }{ \langle \sigma v^2 \rangle_{SSS\to XX} } \approx \left( 1.15\,  \frac{f_{\rm dm}}{g_{\rm dm}} \right)^6 \,.
\end{align}
That means the $3\to2$ rate in the Majorana DM case should be sizable as the vector DM one if $f_{\rm dm} \sim g_{\rm dm}$. We thus conclude that the present study for self-interacting forbidden DM under a cannibally co-decaying phase is also suitable for the Majorana DM Higgs portal model.

\end{document}